\documentclass[12pt]{article}

\usepackage{amsmath}
\usepackage{amsfonts}
\usepackage{graphicx}
\usepackage{caption}
\usepackage{subcaption}

\usepackage{authblk}
\usepackage{amssymb}

\usepackage{epstopdf}
\usepackage{epsfig}
\usepackage{hyperref}

\usepackage{color}

\def\be{\begin{equation}}
\def\ee{\end{equation}}
\def\bea{\begin{eqnarray}}
\def\eea{\end{eqnarray}}

\newcommand{\beq}{\begin{eqnarray}}
\newcommand{\eeq}{\end{eqnarray}}
\newcommand{\ba}{\begin{align}}
\newcommand{\ea}{\end{align}}

\begin{document}

\title{Polar quasinormal modes of the scalarized Einstein-Gauss-Bonnet black holes}

\author[1]{Jose Luis Bl\'azquez-Salcedo \thanks{\href{mailto:jose.blazquez.salcedo@uni-oldenburg.de}{jose.blazquez.salcedo@uni-oldenburg.de}}}

\author[2,3]{Daniela D. Doneva
\thanks{\href{mailto:daniela.doneva@uni-tuebingen.de }{daniela.doneva@uni-tuebingen.de }}}

\author[1]{Sarah Kahlen\thanks{\href{mailto:sarah.kahlen1@uni-oldenburg.de}{sarah.kahlen1@uni-oldenburg.de}}}

\author[1]{Jutta Kunz \thanks{\href{mailto:jutta.kunz@uni-oldenburg.de}{jutta.kunz@uni-oldenburg.de}}}

\author[1,4]{Petya Nedkova \thanks{\href{mailto:pnedkova@phys.uni-sofia.bg}{pnedkova@phys.uni-sofia.bg}}}

\author[2,4,5]{Stoytcho S. Yazadjiev \thanks{\href{mailto:yazad@phys.uni-sofia.bg}{yazad@phys.uni-sofia.bg}}}

\affil[1]{Institute of Physics, Carl von Ossietzky University of Oldenburg, 26111
	Oldenburg, Germany}
\affil[2]{Theoretical Astrophysics, Eberhard Karls University of T\"ubingen, 72076 T\"ubingen, Germany}
\affil[3]{INRNE - Bulgarian Academy of Sciences, 1784  Sofia, Bulgaria}
\affil[4]{Department of Theoretical Physics, Faculty of Physics, Sofia University, 1164 Sofia, Bulgaria}
\affil[5]{Institute of Mathematics and Informatics, Bulgarian Academy of Sciences, Acad. G. Bonchev Street 8, 1113 Sofia, Bulgaria}

\maketitle

\begin{abstract}
We study the polar quasinormal modes of spontaneously scalarized black holes
in Einstein-Gauss-Bonnet theory.
In previous works we showed that a set of nodeless solutions of the fundamental branch
of the model studied in \cite{Doneva:2017bvd} are stable under both radial  \cite{Blazquez-Salcedo:2018jnn}
and axial perturbations  \cite{Blazquez-Salcedo:2020rhf}.
{Here we calculate the polar quasinormal modes and show that this set of solutions is stable against the polar perturbations as well.  Thus for a certain region of the parameter space the scalarized black holes are potentially stable physically interesting objects. The spectrum of the polar quasinormal modes differs both quantitatively and qualitatively from the Schwarzschild one which offers the possibility to test the Gauss-Bonnet theory via the future gravitational wave observations. } 
\end{abstract}

\section{Introduction}

Gravitational wave observations from the merger of compact objects like black holes and neutron stars
provide a new powerful tool to learn about the regime of strong gravity and thus about
General Relativity (GR) and alternative theories of gravity
\cite{TheLIGOScientific:2016src,Abbott:2018lct,LIGOScientific:2019fpa,Will:2014kxa,Berti:2015itd,Barack:2018yly}.
Among the plethora of gravity theories, in particular theories with higher curvature corrections, as motivated
by quantum gravity considerations, have received much attention in recent years.
An attractive class of such theories is Einstein-scalar-Gauss-Bonnet (EsGB) theory,
since they are ghost-free and lead to second order equations of motion
\cite{Horndeski:1974wa,Charmousis:2011bf,Kobayashi:2011nu}.

The coupling function $f(\varphi)$ of the scalar field to the Gauss-Bonnet (GB) term
has a decisive influence on the properties of the resulting EsGB black holes and neutron stars.
Effective low energy string theories feature exponential coupling functions,
with the scalar field representing the dilaton.
These theories possess black holes with scalar hair,
whose properties have been investigated in great detail
\cite{Kanti:1995vq,Torii:1996yi,Guo:2008hf,Pani:2009wy,Pani:2011gy,Kleihaus:2011tg,Ayzenberg:2013wua,Ayzenberg:2014aka,Maselli:2015tta,Kleihaus:2014lba,Kleihaus:2015aje,Blazquez-Salcedo:2016enn,Cunha:2016wzk,Zhang:2017unx,Blazquez-Salcedo:2017txk,Konoplya:2019hml,Zinhailo:2019rwd}.
The GB term allows to circumvent  the no-hair theorems of GR
(see e.g. \cite{Chrusciel:2012jk,Herdeiro:2015waa}),
but these dilatonic theories do not allow for the black hole solutions of GR,
the Schwarzschild solutions and the Kerr solutions.

An interesting rather new development has led to the insight,
that for a whole class of coupling functions $f(\varphi)$ the black hole solutions
of GR remain solutions of the respective EsGB theory,
but become unstable at certain
values of the GB coupling, where branches of scalarized black holes emerge
\cite{Doneva:2017bvd,Antoniou:2017acq,Silva:2017uqg}.
This phenomenon is referred to as curvature-induced spontaneous scalarization,
and is akin to the well-known matter-induced spontaneous scalarization
in scalar-tensor theories in neutron stars \cite{Damour:1993hw}.
By now, a variety of coupling functions has been employed to
study such spontaneously scalarized black holes and their properties
\cite{Doneva:2017bvd,Antoniou:2017acq,Silva:2017uqg,Antoniou:2017hxj,Blazquez-Salcedo:2018jnn,Doneva:2018rou,Minamitsuji:2018xde,Silva:2018qhn,Brihaye:2018grv,Doneva:2019vuh,Myung:2019wvb,Cunha:2019dwb,Macedo:2019sem,Hod:2019pmb,Collodel:2019kkx,Bakopoulos:2020dfg}.

A central question concerning these scalarized black holes is of course their stability.
If physically relevant, these black holes should be stable, at least on astrophysical timescales.
{A coupling function leading to potentially stable spontaneously scalarized black holes
for a wide range of parameters has been proposed in \cite{Doneva:2017bvd}.}
For a fixed value of the GB coupling constant,
there is a critical value $r_{\rm B}$ of the horizon size of the Schwarzschild black hole,
where the fundamental branch of spontaneously scalarized black holes
emerges and continues to exist for all $r_{\rm H}<r_{\rm B}$.
In fact, there is a whole sequence of bifurcation points at smaller values of $r_{\rm H}$,
where radially excited scalarized black holes emerge.
The black holes on the fundamental branch are thermodynamically
{preferred} for the coupling function \cite{Doneva:2017bvd}.
However, their dynamical stability has only partly been investigated up to now.

To determine the dynamical stability, a study of the
linear mode stability of the solutions is a first important step.   
In perturbation theory the modes of the spherically symmetric black holes
can be studied separately according to their parity.
The polar modes have positive parity, while the axial modes have negative parity.
Radial perturbations represent a special subset of the polar modes,
which often reveal already instabilities of the solutions.
We have therefore first considered radial perturbations
for the spontaneously scalarized black holes \cite{Blazquez-Salcedo:2018jnn}.
These have indeed revealed radial instabilities of the Schwarzschild black holes
and of the radially excited scalarized black holes.
However, the spontaneously scalarized black holes on the fundamental branch
do not have unstable radial modes,
when their horizon size $r_{\rm H}$ is in the interval $r_{\rm S1} < r_{\rm H} < r_{\rm B}$.
At the critical value $r_{\rm S1}$ the perturbation equations lose hyperbolicity,
and the employed formalism is no longer well-defined.

To show linear mode stability, we have
next considered axial perturbations \cite{Blazquez-Salcedo:2020rhf}.
Here the scalar field decouples and only gravitational quadrupole
(and higher multipole) modes are present.
Hyperbolicity of the axial perturbation equations is lost as well,
but at a slightly larger value of the horizon size $r_{\rm S2} > r_{\rm S1}$,
for  fixed coupling.
Within the interval $r_{\rm S2} < r_{\rm H} < r_{\rm B}$,
however, the black holes are mode stable with respect to axial perturbations.

In this paper we complete the analysis of linear mode stability
by considering also the non-radial polar modes.
The quadrupole modes now come in two kinds, the scalar-led modes
and the gravitational-led modes. The names imply, that in the
limit of vanishing backgound scalar field, the modes correspond
to the Schwarzschild modes, which are purely scalar
and purely gravitational. For a finite background scalar field, of course
mixing between these channels arises.
The presence of the scalar background field then destroys the
isospectrality present for Schwarzschild black holes.
In addition to the quadrupole modes, the scalarized black holes
possess also dipole modes and scalar modes, which involve both
the scalar field and the gravitational field.
We will see, that all of these modes are stable
in the relevant range $r_{\rm S2} < r_{\rm H} < r_{\rm B}$.

This paper is organized as follows.
In section II we recall the action and the background solutions.
In section III we discuss the equations for the polar perturbations,
including the asymptotic behavior and the numerical method.
We present our numerical results for the polar modes of the
spontaneously scalarized black holes 
in section IV, and we conclude in section V.

\section{Action and background}

The action in Einstein-scalar-Gauss-Bonnet theory is given by
\begin{eqnarray}
S=&&\frac{1}{16\pi}\int d^4x \sqrt{-g}
\Big[R - 2\nabla_\mu \varphi \nabla^\mu \varphi
+ \lambda^2 f(\varphi){\cal R}^2_{GB} \Big] \ ,\label{eq:quadratic}
\end{eqnarray}
where the spacetime metric is $g_{\mu\nu}$ with Ricci scalar $R$,
$\varphi$ is the scalar field with coupling function $f(\varphi)$, and $\lambda$  is the GB coupling constant
with dimension of $length$.
The GB invariant ${\cal R}^2_{GB}$ is defined as
${\cal R}^2_{GB}=R^2 - 4 R_{\mu\nu} R^{\mu\nu}
+ R_{\mu\nu\alpha\beta}R^{\mu\nu\alpha\beta}$
with Ricci tensor $R_{\mu\nu}$
and Riemann tensor $R_{\mu\nu\alpha\beta}$.

The field equations that result from this action are
\begin{eqnarray}\label{FE1}
&&R_{\mu\nu}- \frac{1}{2}R g_{\mu\nu} + \Gamma_{\mu\nu}= 2\nabla_\mu\varphi\nabla_\nu\varphi -  g_{\mu\nu} \nabla_\alpha\varphi \nabla^\alpha\varphi \ ,\\
&&\nabla_\alpha\nabla^\alpha\varphi=  -  \frac{\lambda^2}{4} \frac{df(\varphi)}{d\varphi} {\cal R}^2_{GB} \ , \label{FE2}
\end{eqnarray}
where
\begin{eqnarray}
\Gamma_{\mu\nu}&=& - R(\nabla_\mu\Psi_{\nu} + \nabla_\nu\Psi_{\mu} ) - 4\nabla^\alpha\Psi_{\alpha}\left(R_{\mu\nu} - \frac{1}{2}R g_{\mu\nu}\right) +
4R_{\mu\alpha}\nabla^\alpha\Psi_{\nu} + 4R_{\nu\alpha}\nabla^\alpha\Psi_{\mu} \nonumber \\
&& - 4 g_{\mu\nu} R^{\alpha\beta}\nabla_\alpha\Psi_{\beta}
+ \,  4 R^{\beta}_{\;\mu\alpha\nu}\nabla^\alpha\Psi_{\beta} \ , \\
\Psi_{\mu}&=& \lambda^2 \frac{df(\varphi)}{d\varphi}\nabla_\mu\varphi \ .
\end{eqnarray}

In the present paper we will use the coupling function
introduced in \cite{Doneva:2017bvd}
\begin{equation} \label{eq:coupling_function}
f(\varphi)=  \frac{1}{12} \left(1- e^{-6\varphi^2}\right) \ .
\end{equation}
Note that the coupling function $f(\varphi)$ satisfies
the conditions
$\frac{df}{d\varphi}(0)=0$ and $b^2=\frac{d^2f}{d\varphi^2}(0)>0$,
and hence we have curvature induced spontaneous scalarization
of black holes in the theory,
when we fix the cosmological value of the scalar field to zero,
$ \varphi_{\infty}=0$
\cite{Doneva:2017bvd,Antoniou:2017acq,Silva:2017uqg}.

\begin{figure}
	\centering
	\includegraphics[width=0.34\linewidth,angle=-90]{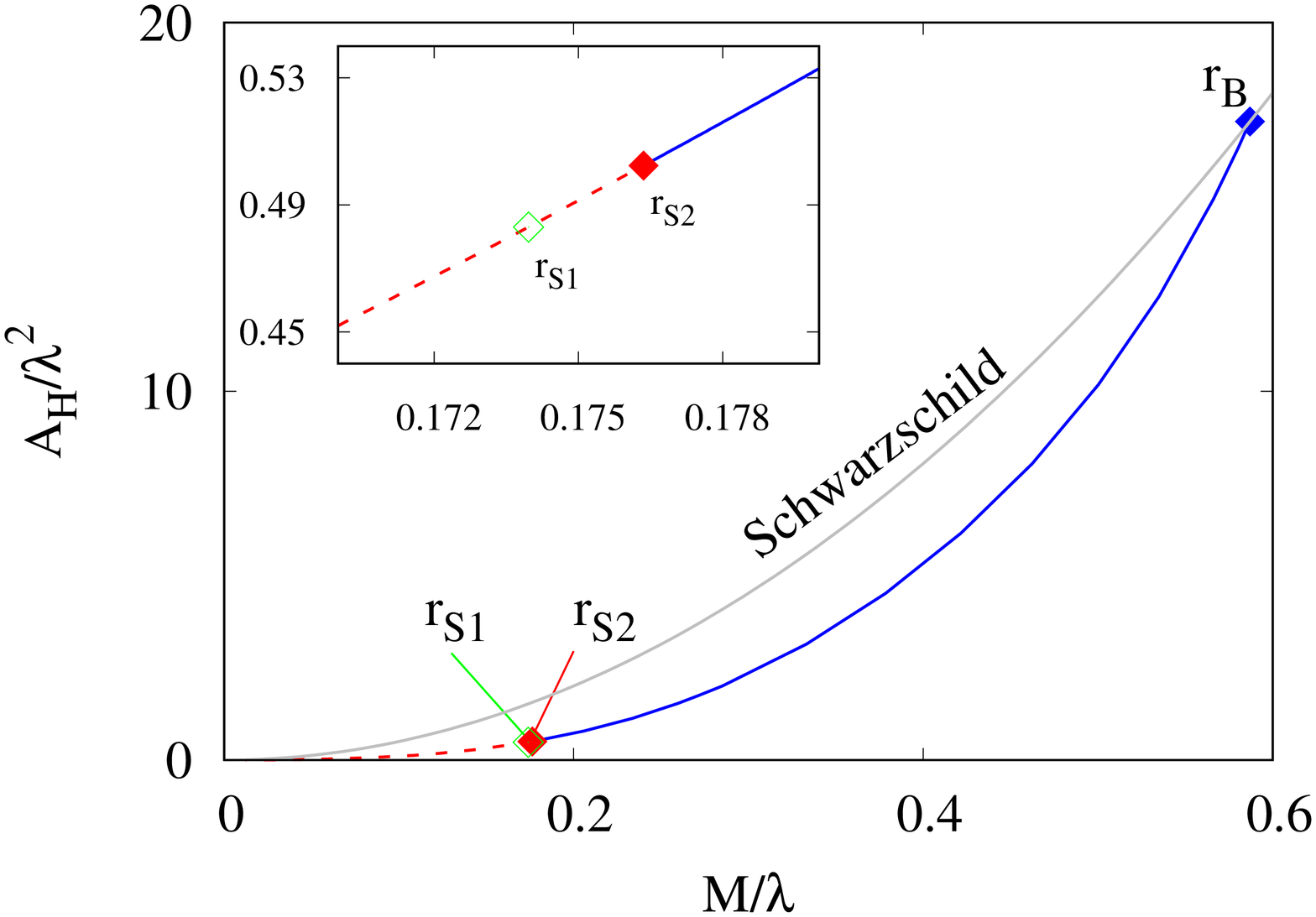}
	\includegraphics[width=0.34\linewidth,angle=-90]{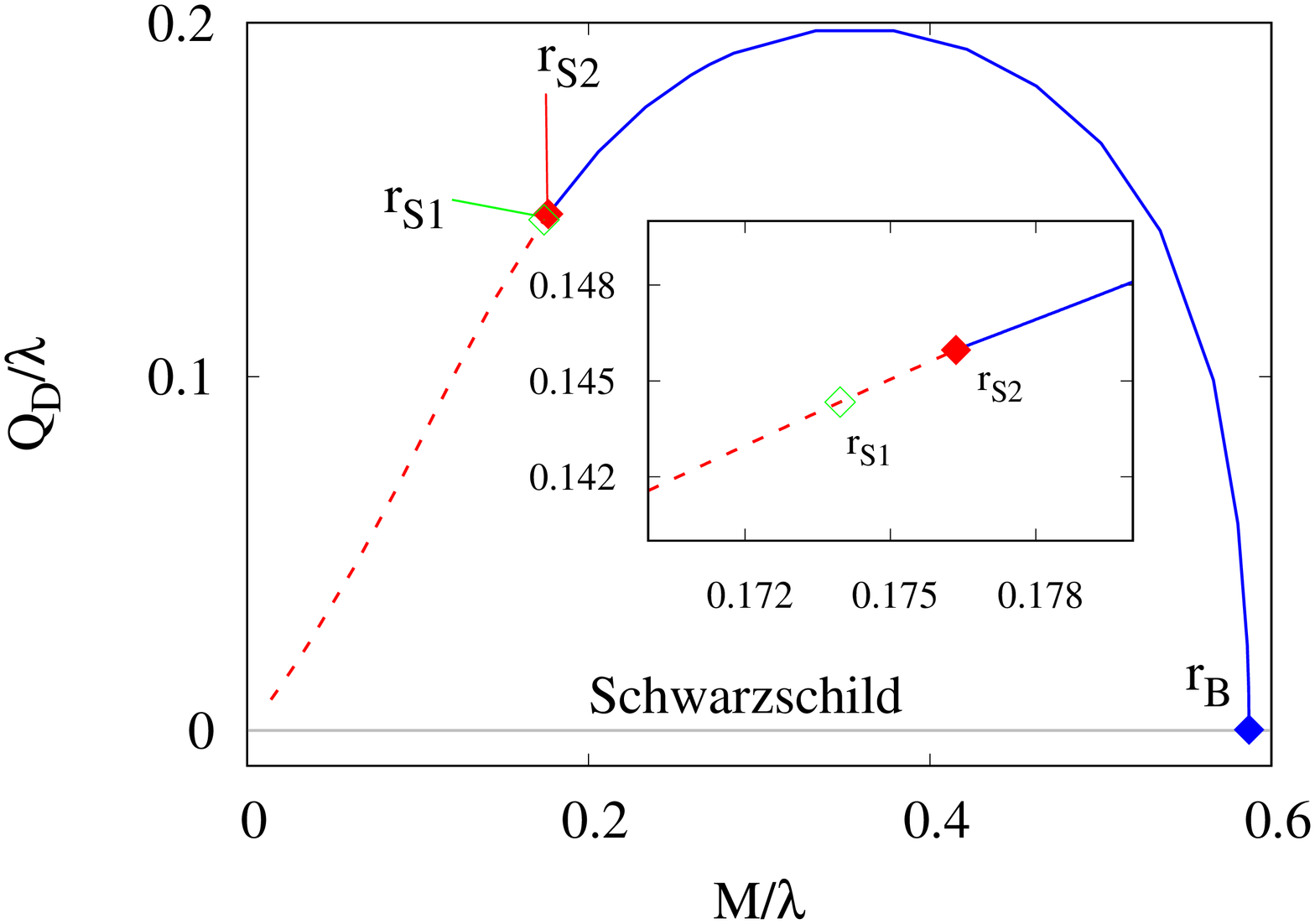}\\[2ex]
	\caption{(\textit{left}) Scaled horizon area $A_{\rm H}/\lambda^2$
vs. scaled total mass $M/\lambda$, and
(\textit{right}) Scaled scalar charge $Q_{\rm D}/\lambda$ vs. scaled total mass
for {the fundamental branch of EsGB black holes }(blue and red) and the Schwarzschild black holes (grey).
The bifurcation point $r_{\rm B}$ is marked in blue,
the points $r_{\rm S1}$ and $r_{\rm S2}$, where hyperbolicity is lost
for the radial and axial perturbation equations, are marked in green and red, respectively.}
	\label{fig:static}
\end{figure}

Spherically symmetric solutions
of the field equations can be obtained with the Ansatz for the metric
\begin{eqnarray}\label{eq:metric_BG}
ds^2= - f(r)dt^2 + \frac{1}{1-\frac{2m(r)}{r}} dr^2
+ r^2 (d\theta^2 + \sin^2\theta d\phi^2 ) \ ,
\end{eqnarray}
where the metric functions $f(r)$ and $m(r)$ depend only on the radial coordinate,
and the scalar field $\varphi(r)$ is likewise only a function of $r$.
By solving the set of field equations for these functions
subject to the conditions of asymptotic flatness and regularity at and outside the horizon,
the domain of existence of spontaneously scalarized black holes solutions has been
mapped out in \cite{Doneva:2017bvd}.
In fact, one can work with $\lambda=1$ without loss of generality,
since all dimensionful quantities can be rescaled with respect to $\lambda$.
The black holes can then be parametrized by the value
of the horizon radius $r_{\rm H}$.

Here we focus only on the fundamental branch of the scalarized black holes,
since this should be the physically most relevant branch of solutions.
This branch is shown together with the Schwarzschild branch
in Fig.~\ref{fig:static},
where the scaled horizon area $A_{\rm H}/\lambda^2$ (left)
and the scaled scalar charge $Q_{\rm D}/\lambda$ (right) are exhibited
as functions of the scaled mass $M/\lambda$.
Note, that the scalar charge $Q_{\rm D}$ is defined as the coefficient of the dominant
term in the asymptotic expansion of the scalar field.
The fundamental branch bifurcates from the Schwarzschild branch
at the horizon radius $r_{\rm B}=1.173944$ and continues to exist for all
radii $r_{\rm H}<r_{\rm B}$.
The fundamental branch is marked in blue and red in the figures,
while the Schwarzschild branch is marked in grey.

Analysis of the radial perturbation equations for the solutions on the fundamental
branch shows, that hyperbolicity of the equations is lost at
a critical horizon radius $r_{\rm S1}=0.191605$.
However, in the interval $r_{\rm S1} < r_{\rm H}< r_{\rm B}$
all solutions are stable with respect to radial perturbations \cite{Blazquez-Salcedo:2018jnn} .
Similarly, analysis of the axial perturbation equations
reveals loss of hyperbolicity at $r_{\rm S2}=0.19994$,
but stability with respect to axial perturbations
in the interval $r_{\rm S2} < r_{\rm H}< r_{\rm B}$ \cite{Blazquez-Salcedo:2020rhf}.
The critical points $r_{\rm S1}$ and $r_{\rm S2}$ ($r_{\rm S1} < r_{\rm S2}$)
are marked in Fig.~\ref{fig:static} in green and red, respectively.

We conclude, that there is a potentially physically interesting stable range
$r_{\rm S2} < r_{\rm H}< r_{\rm B}$ of scalarized black holes.
To show mode stability of these black holes we still have to
investigate their non-radial polar modes.
This will be done in the following {sections}.

\section{Polar Perturbations}

\subsection{Ansatz}

Introducing a perturbation control parameter $\epsilon$, and the spherical harmonics $Y_{lm}(\theta,\phi)$ with angular numbers $l$ and $m$, we  adopt for the polar perturbation {of the metric the following Ansatz in the Regge-Wheeler gauge}
\cite{Regge:1957td}
\begin{eqnarray}
\label{eq:pert_metric}
ds^2=-f(r)\left[1+\epsilon e^{-i\omega t}H_0(r)Y_{lm}(\theta,\phi)\right]dt^2+\frac{1}{1-\frac{2m(r)}{r}}\left[1+\epsilon e^{-i\omega t}{H_2}(r) Y_{lm}(\theta,\phi)\right]dr^2 \nonumber \\ +r^2\left[1+\epsilon e^{-i\omega t}{T}(r) Y_{lm}(\theta,\phi)\right](d\theta^2 + \sin^2\theta d\phi^2 ) - 2\epsilon  e^{-i\omega t}H_1(r)Y_{lm}(\theta,\phi)drdt \ ,
\end{eqnarray}
and for the scalar field
\begin{eqnarray}
\label{eq:pert_scalar}
\varphi= \varphi_0(r) \left[1+\epsilon e^{-i\omega t}\varphi_1(r)Y_{lm}(\theta,\phi)\right] \ .
\end{eqnarray}

The resulting set of equations is given in the Appendix. {Due to the spherical symmetry of the background solution the radial perturbation equations do not depend on the value of $m$.}
{They can be reduced to} a second order differential equation for the scalar field perturbation function $\varphi_1$ coupled to two first order differential equations for the metric perturbation functions $H_1$ and $T$. {After solving this system the remaining functions $H_0$ and $H_2$ can be obtained by means of two relations, which represent them by means of the rest of the perturbation fields.}

The coupled system can be expressed in the schematic first order form
\be
\frac{d}{dr}{{\Psi}}+{V}{\Psi}= 0
\label{eq:perturbations}
\ee
where ${\Psi}$ is a column vector with components $(H_1,T,\varphi_1,\varphi_1')$.
The matrix ${V}$ contains the coupling between the background functions and the perturbation functions.
The set of coupled first order equations can also be cast
into a set of coupled second-order Schr\"odinger-like equations.

Let us address here what happens in the Schwarzschild limit.
At the bifurcation point $r_{\rm H}=r_{\rm B}$, the scalar field of the solution vanishes, and the exterior of the black hole is described by the Schwarzschild solution, with $f(r)=1-\frac{2M}{r}$, $m(r)=M$ and $\phi_0(r)=0$. In practice, the polar perturbations decouple into two sets of second order equations: one set for $(H_1,T)$, that can be written like the Zerilli equation using the standard definitions we shall not repeat here. The second set for $(\varphi_1,\varphi_1')$ results in the following equation for the scalar field perturbations
\begin{eqnarray}
\frac{d^2 \varphi_1}{dr^2}=-\frac{2(r-M)}{r\left(r-2M\right)}\frac{d \varphi_1}{dr}+\left[
\frac{l(l+1)}{r(r-2M)}
-\frac{12\lambda^2M^2}{r^5(r-2M)}
-\frac{r^2\omega^2}{(r-2M)^2}
\right] \varphi_1 \, .
\end{eqnarray}
Note that this equation is not exactly the one for a minimally coupled scalar field, since it has a source term proportional to the Gauss-Bonnet invariant and the coupling constant $\lambda^2$. In practice, this means that the scalar modes at the bifurcation point will be shifted with respect to the standard scalar modes of Schwarzschild that one obtains in the $\lambda=0$ coupling limit.

\subsection{Asymptotic behaviour and numerical method}

For the study of the asymptotic behaviour of the perturbations, we introduce the tortoise coordinate
\begin{eqnarray}
\frac{dR^*}{dr} = \frac{1}{\sqrt{f(r)\left(1-\frac{2m(r)}{r}\right)}} ,
\end{eqnarray}
and express the spatial derivatives in terms of this coordinate.
The quasinormal modes are then obtained by solving the set of equations
\eqref{eq:perturbations} subject to the proper boundary conditions,
which correspond to purely outgoing waves at infinity
and purely ingoing waves at the black hole horizon
\be \Psi \propto
\left\{ \begin{array}{ll} e^{-i\omega (t+R_*)}\,,
&r\to r_{\rm H} \,,\\
e^{-i\omega (t-R_*)}\,,
&r\to\infty\,. \end{array} \right.
\ee

The far-field behavior of the perturbations is determined by the expansions
\begin{eqnarray}
T &=& e^{i\omega R^*} \left[ A_g^+ \left(1 + \frac{1}{r^2}\left(\frac{Q_D^2}{2}-\frac{3iM}{2\omega}+\frac{l(l-1)(l+2)}{8\omega^2}\right)+  O(r^{-3}) \right) + \right. \nonumber \\
&&\left. + A_s^+\left(\frac{-2i Q_D}{\omega r^3} +  \frac{Q_D(l^2+l-3)}{\omega^2 r^4} +  O(r^{-5}) \right) \right] , \\
H_1 &=& r \omega e^{i\omega R^*} \left[ A_g^+ \left(1 + \frac{1}{r}\left(2M+\frac{i(l-1)(l+2)}{2\omega}\right)+  O(r^{-2}) \right) + \right. \nonumber \\
&&\left. + A_s^+\left(\frac{-Q_D}{\omega^2 r^4} +  \frac{Q_D}{r^5}\left(\frac{24M\lambda^2}{5}+\frac{2M}{5\omega^2}-\frac{i(l-2)(l+3)}{2\omega^3}\right) +  O(r^{-6}) \right) \right] ,
\end{eqnarray}
\begin{eqnarray}
 \varphi_1 &=& \frac{1}{r} e^{i\omega R^*} \left[ A_g^+ \left(\frac{-MQ_D}{r}-\frac{Q_D}{r^2}\left(2M^2+\frac{iM(l+2)(l-1)}{2\omega}\right)+  O(r^{-3}) \right) + \right. \nonumber \\
 &&\left. + A_s^+\left(1+\frac{il(l+1)}{2\omega r} + O(r^{-2}) \right) \right] ,
\end{eqnarray}
where $A_s^+$ and $A_g^+$ are the scalar  and
space-time perturbation amplitudes, respectively, characterizing the expansions.
The far-field expansion is written in terms of the global parameters of the background solution, the total mass $M$ and the scalar charge $Q_D$.
The near horizon behavior on the other hand is given by the expansions
\begin{eqnarray}
T &=&  e^{-i\omega R^*} \left[ A_g^- + O(r-r_{\rm H}) \right] \,  \\
\varphi_1 &=& \frac{1}{r} e^{-i\omega R^*} \left[ A_s^- + O(r-r_{\rm H}) \right]\, \\
H_1 &=& \frac{1}{r-r_{\rm H}} e^{-i\omega R^*} \left[ A_s^-\left(\frac{-4\lambda^2 \varphi_{\rm H}}{r_{\rm H}^3e^{6 \varphi_{\rm H}^2}} +  O(r-r_{\rm H}) \right) +\right.  \\
 &&\hskip -1cm \left. + A_g^- \left(\frac{4\omega^2r_{\rm H}+2i\omega\sqrt{f'_{\rm H} r_{\rm H}(1-2m'_{\rm H})(l^2+l+1)}-f'_{\rm H}(1-2m'_{\rm H})l(l+1)}{r_{\rm H}^2(f'_{\rm H}(1-2m'_{\rm H})+4r_{\rm H}\omega^2)} +  O(r-r_{\rm H}) \right) \right] \, \nonumber
\end{eqnarray}
where $A_s^-$ and $A_g^-$ are again the scalar and space-time perturbation amplitudes.
In the near horizon expansion we have defined the horizon parameters $ \varphi_{\rm H}= \varphi_0(r_{\rm H})$, $ \varphi'_{\rm H}=\frac{d \varphi_0}{dr}|_{r_{\rm H}}=$, $m'_{\rm H}=\frac{dm}{dr}|_{r_{\rm H}}$ and $f'_{\rm H}=\frac{df}{dr}|_{r_{\rm H}}$, which can be obtained from the background solution. Not all of these background near-horizon parameters are free, since in particular the horizon expansion at zero order imposes the relations $(2\lambda^2 \varphi_{\rm H} \varphi'_{\rm H}+r_{\rm H}e^{6 \varphi_{\rm H}^2})m'_{\rm H}=\lambda^2 \varphi_{\rm H} \varphi'_{\rm H}$ and $\lambda^2r_{\rm H} \varphi_{\rm H} \varphi'_{\rm H}=\sqrt{r_{\rm H}^4e^{12 \varphi_{\rm H}^2}-24\lambda^2 \varphi_{\rm H}}-r_{\rm H}e^{6 \varphi_{\rm H}^2}$.

The eigenvalue $\omega=\omega_R+i\omega_I$ consists of a real part $\omega_R$,
corresponding to the frequency of the mode, and of an imaginary part $\omega_R$,
corresponding to the damping rate, as long as it is positive.
A negative imaginary part, however, implies an instability.
Thus a criterion for linearized mode stability of the black holes is
that all quasinormal mode frequencies have a positive imaginary part.

To calculate the quasinormal mode frequencies, we employ a shooting method,
that is generating solutions that satisfy the near-horizon expansion, and solutions that satisfy the far-field expansion, with different values of the scalar and space-time amplitudes $A_s^\pm$ and $A_g^\pm$.
Subsequently we match them at an intermediate point.
Continuity of the functions and their derivatives then determine the values of $\omega$,
that define the quasinormal modes. We always calibrate and cross-check the numerics with the different limits that we can continuously study (in this case the GR modes when $\lambda=0$, or the Schwarzschild modes when $ \varphi_0(r)=0$).
Further details on the numerical procedure can, for instance, be found in \cite{Blazquez-Salcedo:2018pxo}.

\section{Results}

In the following, we will present our results
for the polar quasinormal modes of the spontaneously scalarized EsGB black holes
on the fundamental branch in the physically interesting range
$r_{\rm S2} < r_{\rm H}< r_{\rm B}$,
concluding the demonstration of their mode stability in this range.
We note, that we have extended the calculations of the presented modes
into the range $ r_{\rm H} < r_{\rm S2}$,
without finding any pecularity there
indicating a further instability or another breakdown of hyperbolicity.

We will discuss the modes in the order of their angular number $l$, comprising
the monopole case $l=0$, the dipole case $l=1$, and the quadrupole case $l=2$.
In the latter case we have to distinguish between two families of modes \cite{Blazquez-Salcedo:2016enn}:
\begin{itemize}
\item[i.]
Modes connected with purely gravitational perturbations
of the Schwarzschild solution are referred to as gravitational-led (grav-led) modes.
These have dominant amplitude $A_g^{\pm}$.
\item[ii.]
Modes connected with purely scalar perturbations
of the Schwarzschild solution are referred to as scalar-led modes.
These have dominant amplitude $A_s^{\pm}$.
\end{itemize}
We will also demonstrate the breaking of isospectrality.

\subsection{$l=0$}

\begin{figure}
\centering
	\includegraphics[width=0.34\linewidth,angle=-90]{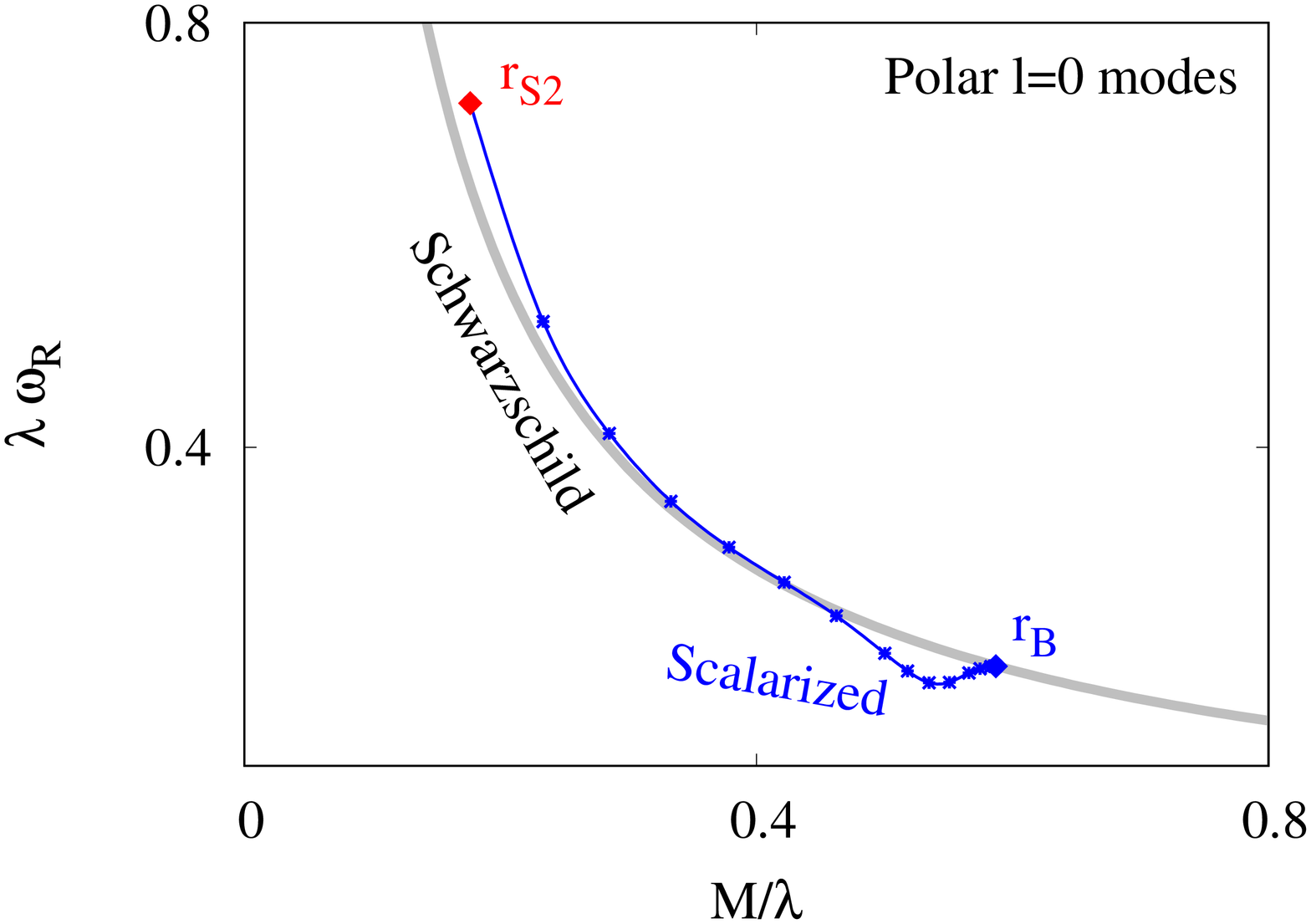}
	\includegraphics[width=0.34\linewidth,angle=-90]{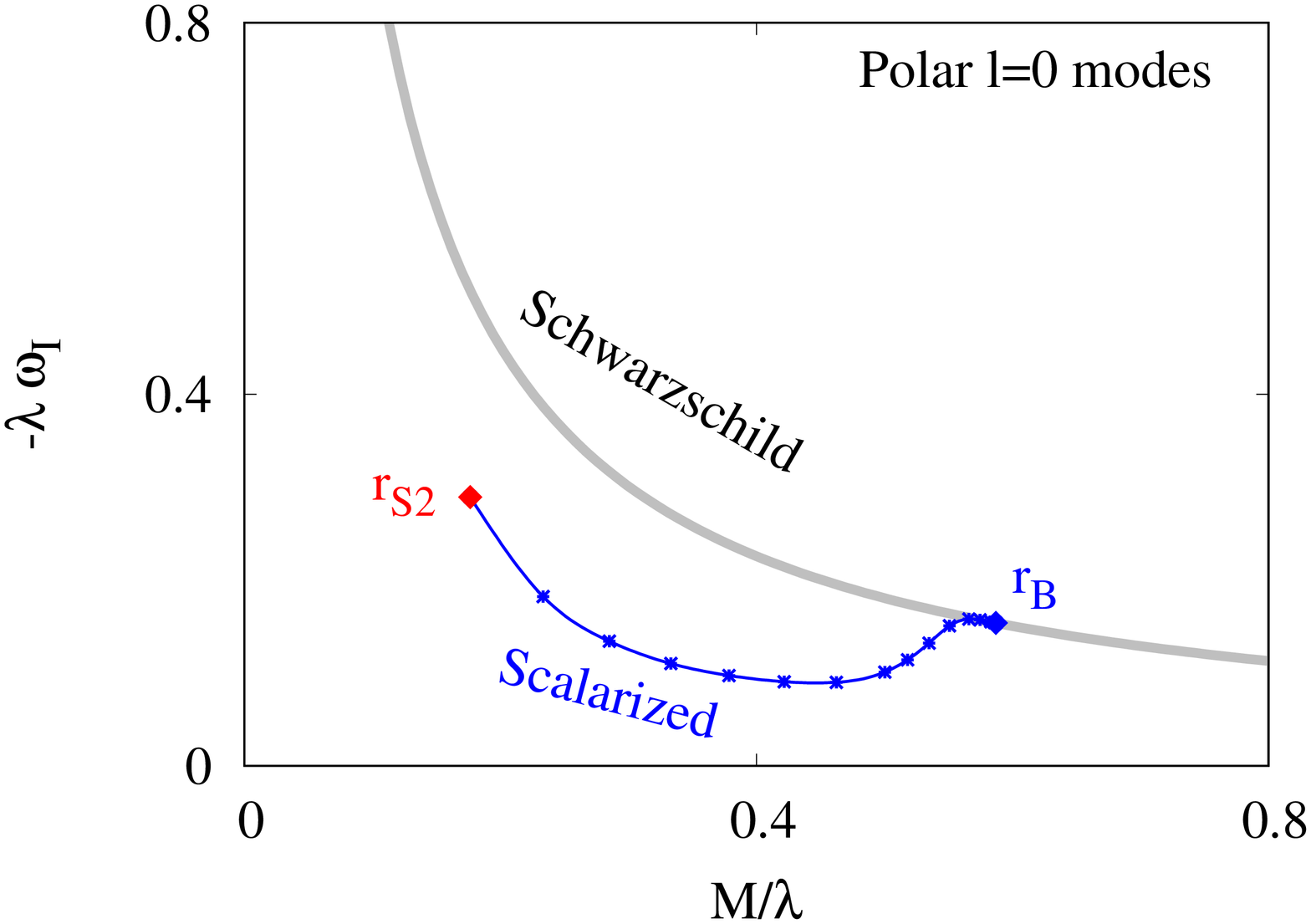}
	\caption{
Scaled polar $l=0$ eigenvalue $\lambda \omega$
vs. scaled total mass $M/\lambda$: real part/frequency $\omega_R$
(\textit{left}) and imaginary part/inverse damping time $\omega_I$ (\textit{right}),
depicted in the range $r_{\rm S2} < r_{\rm H}< r_{\rm B}$,
where hyperbolicity is lost at $r_{\rm S2}$, and
$r_{\rm B}$ is the bifurcation point from the Schwarzschild solution.
For comparison also the Schwarzschild mode is shown (solid grey).}
	\label{fig:plotl0}
\end{figure}
\begin{figure}
	\centering
	\includegraphics[width=0.34\linewidth,angle=-90]{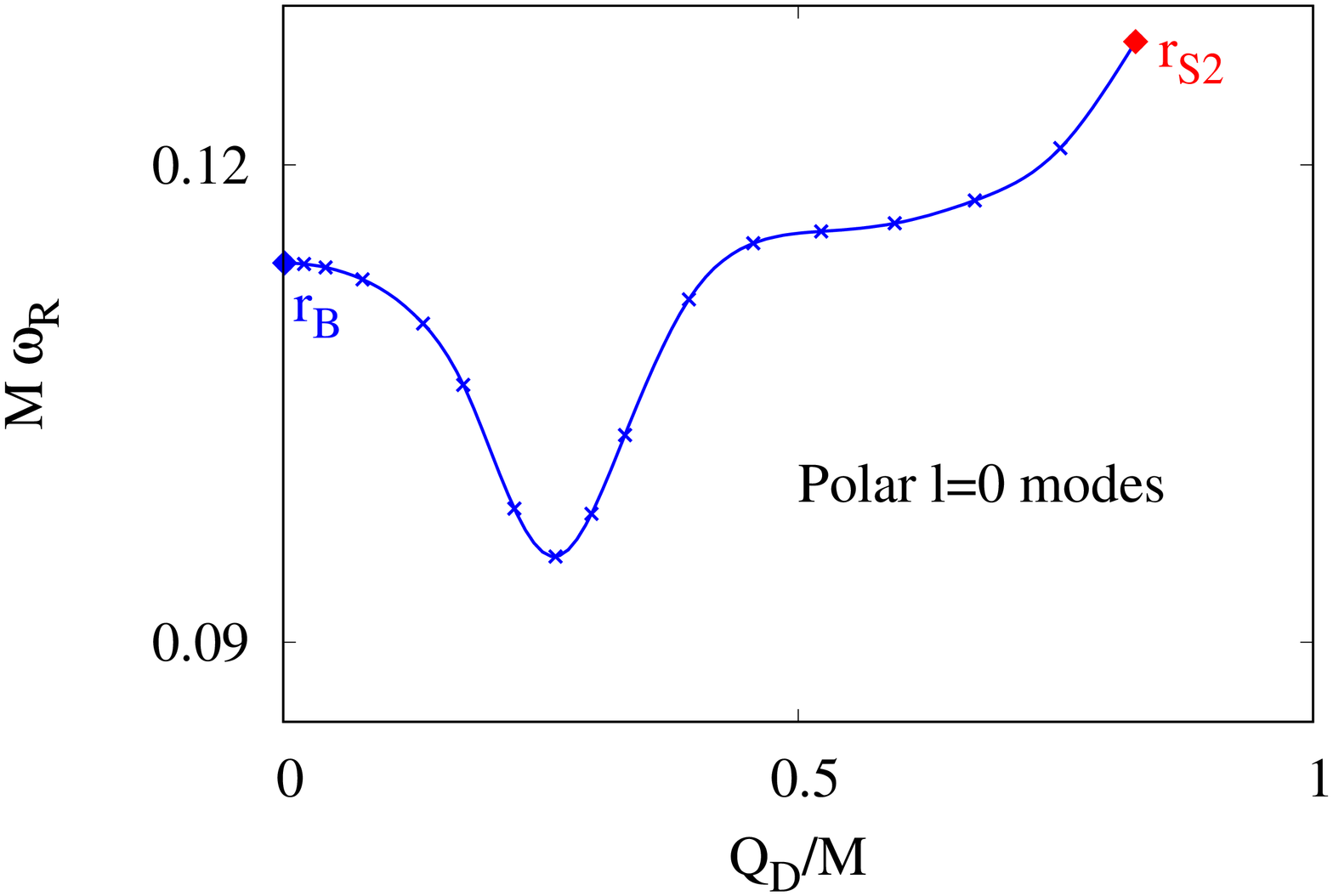}
	\includegraphics[width=0.34\linewidth,angle=-90]{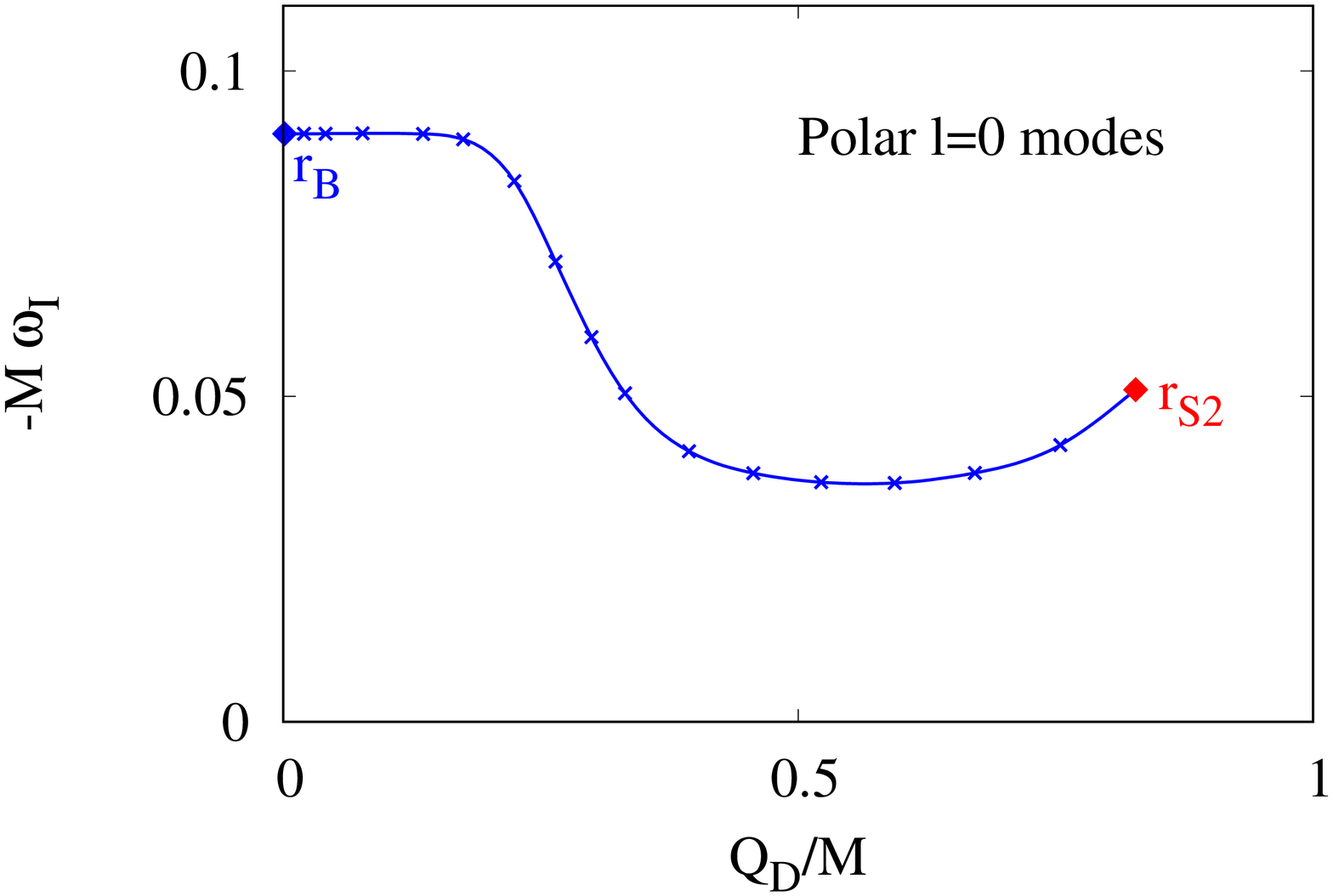}
	\caption{
Scaled polar $l=0$ eigenvalue $M \omega$
vs. scalar charge $Q_D/M$: real part/frequency $\omega_R$
(\textit{left}) and imaginary part/inverse damping time $\omega_I$ (\textit{right}),
depicted in the range $r_{\rm S2} < r_{\rm H}< r_{\rm B}$.
Hyperbolicity is lost at $r_H=r_{\rm S2}$, the maximum value of $Q_D/M$ shown.
$Q_D/M=0$ corresponds to the bifurcation point $r_{\rm B}$
from the Schwarzschild solution.}
	\label{fig:plotl0_Qd}
\end{figure}

Let us now consider the results for the monopole ($l=0$) modes.
They are obtained by starting at the bifurcation point $r_{\rm H} = r_{\rm B}$,
where a corresponding mode of the Schwarzschild solution
refers to an independent scalar field in  the Schwarzschild background.
This mode is given by $M\omega=0.114-i0.090$.
From $ r_{\rm B}$ the mode is then tracked along the fundamental branch.
The $l=0$ mode is exhibited in Fig.~\ref{fig:plotl0},
where the scaled eigenvalue $\lambda \omega$ is shown
versus the scaled total mass $M/\lambda$.
The left plot exhibits the real part $\omega_R$,
representing the frequency,
and the right plot exhibits the imaginary part $\omega_I$,
representing the inverse damping time.
For comparison also the Schwarzschild mode is shown (solid grey).
We observe that the scaled frequency of the scalarized black holes
of a given mass follows closely the corresponding frequency
of the Schwarzschild black holes.
The damping rate of the scalarized black holes
is always smaller than its Schwarzschild counterpart,
thus the damping time is larger for the scalarized solutions.

In Fig.~\ref{fig:plotl0_Qd} we illustrate the dependence of the scaled eigenvalue
$M \omega$ on the scaled scalar charge $Q_D/M$ of the black holes.
The corresponding Schwarzschild values are read off at $Q_D/M=0$.
The scaled frequency varies by less than 20\%, whereas
the damping rate decreases strongly for larger values of the scaled scalar charge.
We emphasize that we do not find unstable modes.

\subsection{$l=1$}

\begin{figure}
	\centering
	\includegraphics[width=0.34\linewidth,angle=-90]{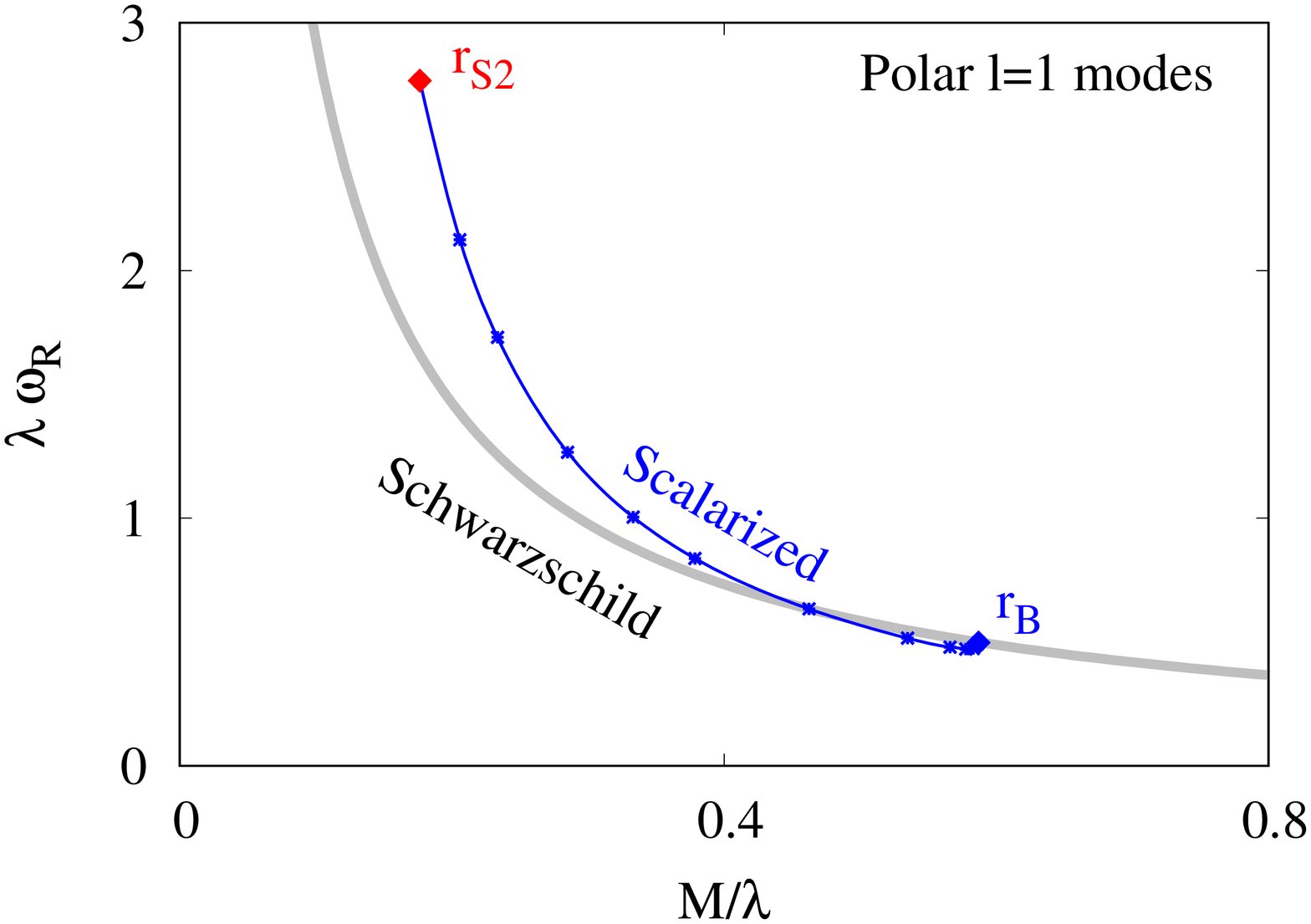}
	\includegraphics[width=0.34\linewidth,angle=-90]{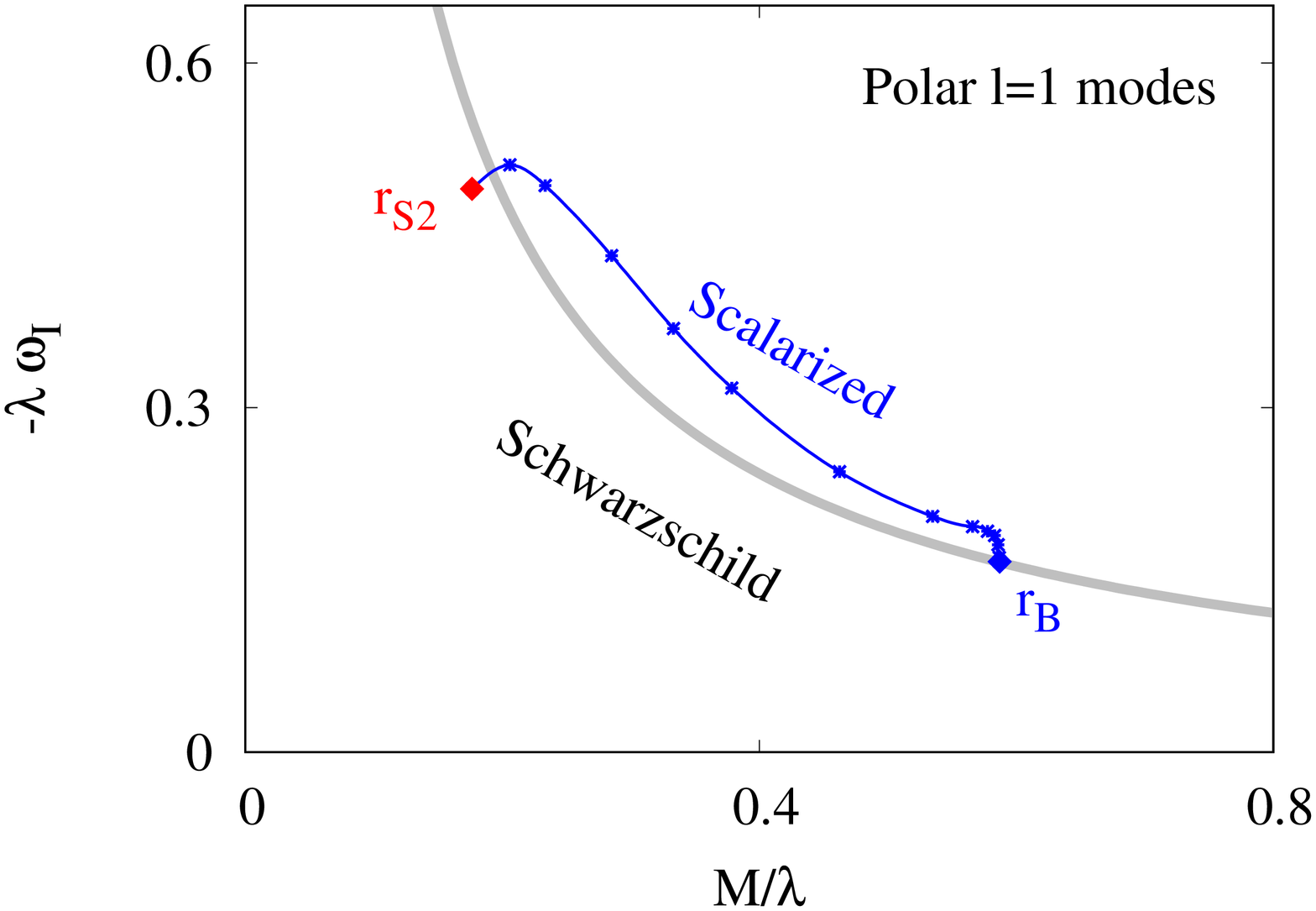}
	\caption{
Scaled polar $l=1$ eigenvalue $\lambda \omega$
vs. scaled total mass $M/\lambda$: real part/frequency $\omega_R$
(\textit{left}) and imaginary part/inverse damping time $\omega_I$ (\textit{right}),
depicted in the range $r_{\rm S2} < r_{\rm H}< r_{\rm B}$,
where hyperbolicity is lost at $r_{\rm S2}$, and
$r_{\rm B}$ is the bifurcation point from the Schwarzschild solution.
For comparison also the Schwarzschild mode is shown (solid grey).}
	\label{fig:plotl1}
\end{figure}
\begin{figure}
	\centering
	\includegraphics[width=0.34\linewidth,angle=-90]{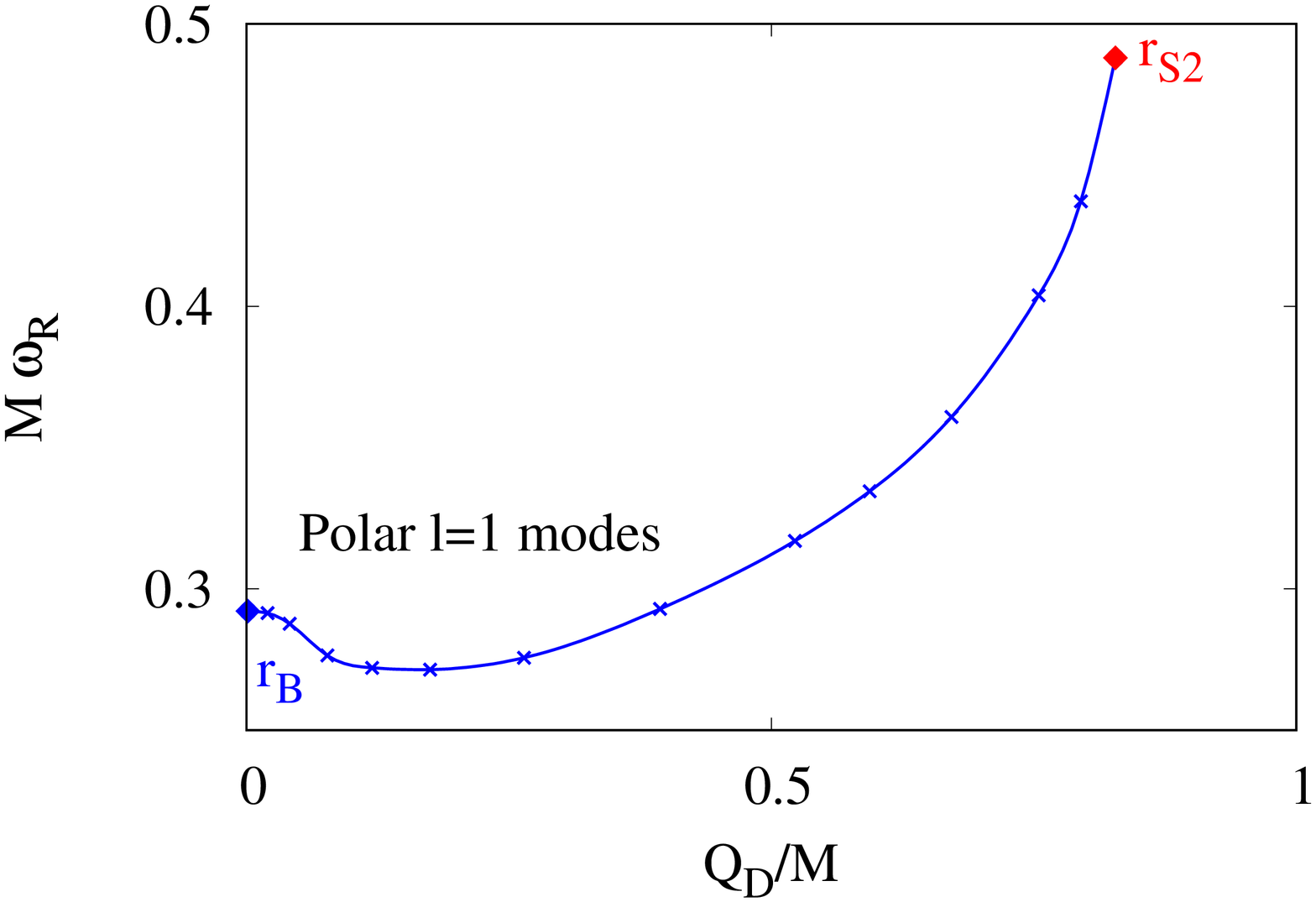}
	\includegraphics[width=0.34\linewidth,angle=-90]{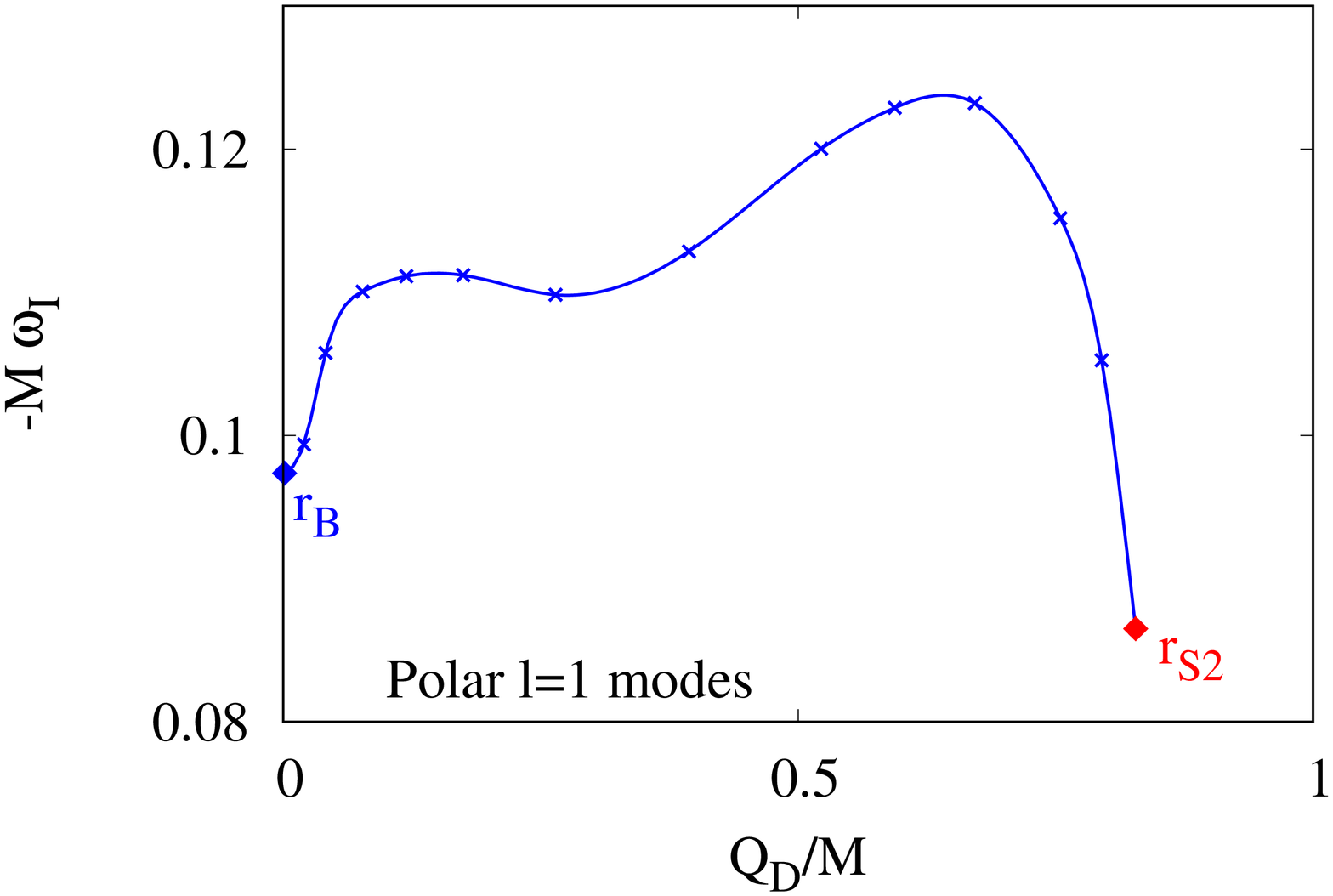}
	\caption{
Scaled polar $l=1$ eigenvalue $M \omega$
vs. scalar charge $Q_D/M$: real part/frequency $\omega_R$
(\textit{left}) and imaginary part/inverse damping time $\omega_I$ (\textit{right}),
depicted in the range $r_{\rm S2} < r_{\rm H}< r_{\rm B}$.
Hyperbolicity is lost at $r_H=r_{\rm S2}$, the maximum value of $Q_D/M$ shown.
$Q_D/M=0$ corresponds to the bifurcation point $r_{\rm B}$
from the Schwarzschild solution.}
	\label{fig:plotl1_Qd}
\end{figure}

To obtain the dipole ($l=1$) modes,  we proceed analogously to the $l=0$ case.
We start again at the bifurcation point $r_{\rm H} = r_{\rm B}$,
with the corresponding mode of the Schwarzschild solution
$M\omega=0.293-i0.097$,
and then track the mode along the fundamental branch.
The $l=1$ mode is exhibited in Fig.~\ref{fig:plotl1},
where again $\lambda \omega$ is shown versus $M/\lambda$,
with frequency $\omega_R$ on the left side and damping rate $\omega_I$ on the right.
For comparison again the Schwarzschild mode is shown (solid grey).
We observe that the  frequency and the damping time of the scalarized black holes
follow more or less the corresponding frequency and damping  time
of the Schwarzschild black holes, being mostly slightly larger than
their Schwarzschild counterparts.
When scaled with the mass and considered as a function of the scalar charge,
the deviations from the Schwarzschild values become more pronounced
as seen in Fig.~\ref{fig:plotl1_Qd}.
Again, we emphasize that we do not find unstable modes.

\subsection{$l=2$}

\begin{figure}
	\centering
	\includegraphics[width=0.34\linewidth,angle=-90]{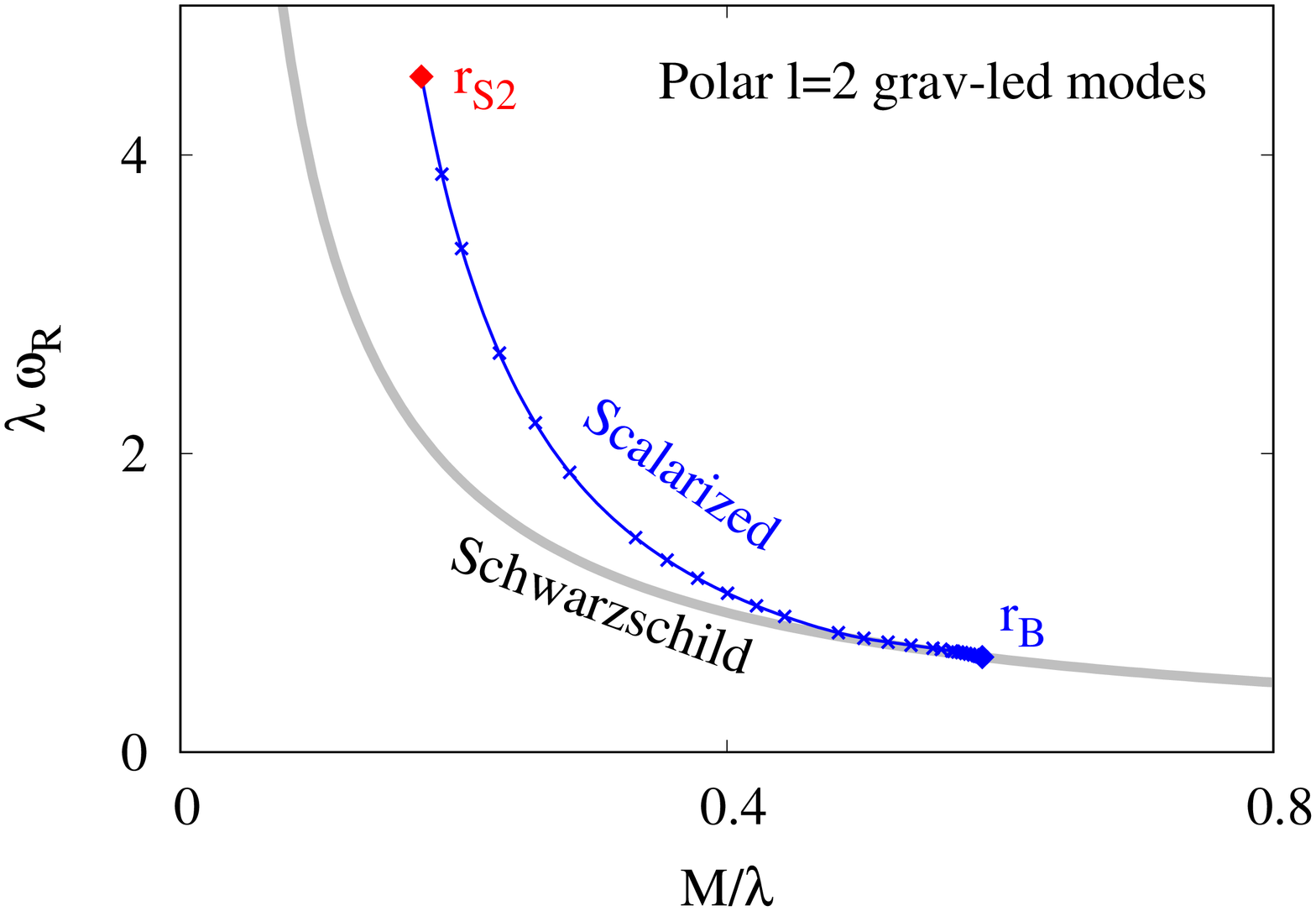}
	\includegraphics[width=0.34\linewidth,angle=-90]{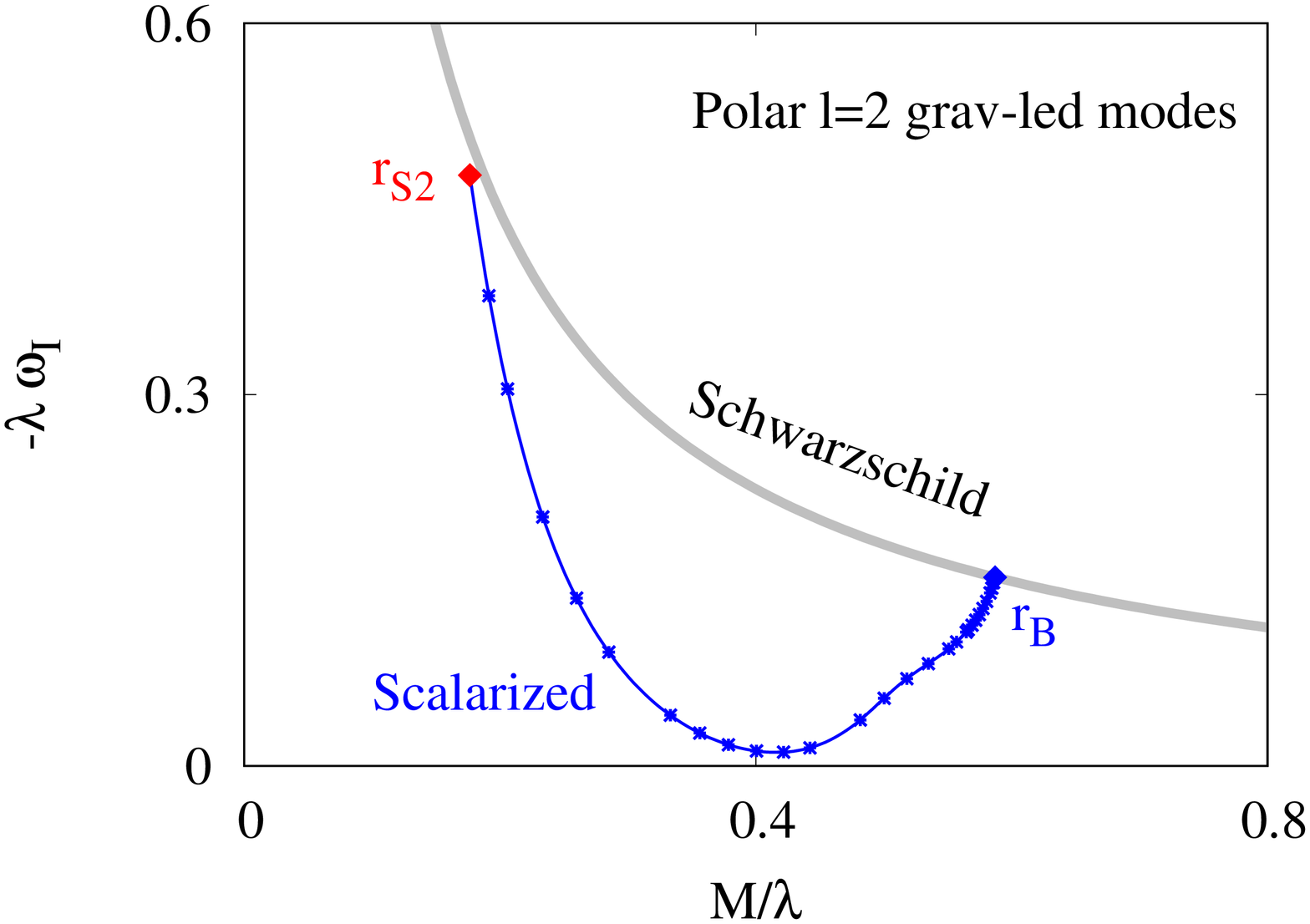}
	\caption{
Scaled polar grav-led $l=2$ eigenvalue $\lambda \omega$
vs. scaled total mass $M/\lambda$: real part/frequency $\omega_R$
(\textit{left}) and imaginary part/inverse damping time $\omega_I$ (\textit{right}),
depicted in the range $r_{\rm S2} < r_{\rm H}< r_{\rm B}$,
where hyperbolicity is lost at $r_{\rm S2}$, and
$r_{\rm B}$ is the bifurcation point from the Schwarzschild solution.
For comparison also the Schwarzschild mode is shown (solid grey).}
	\label{fig:plotl2_grav}
\end{figure}
\begin{figure}
	\centering
	\includegraphics[width=0.34\linewidth,angle=-90]{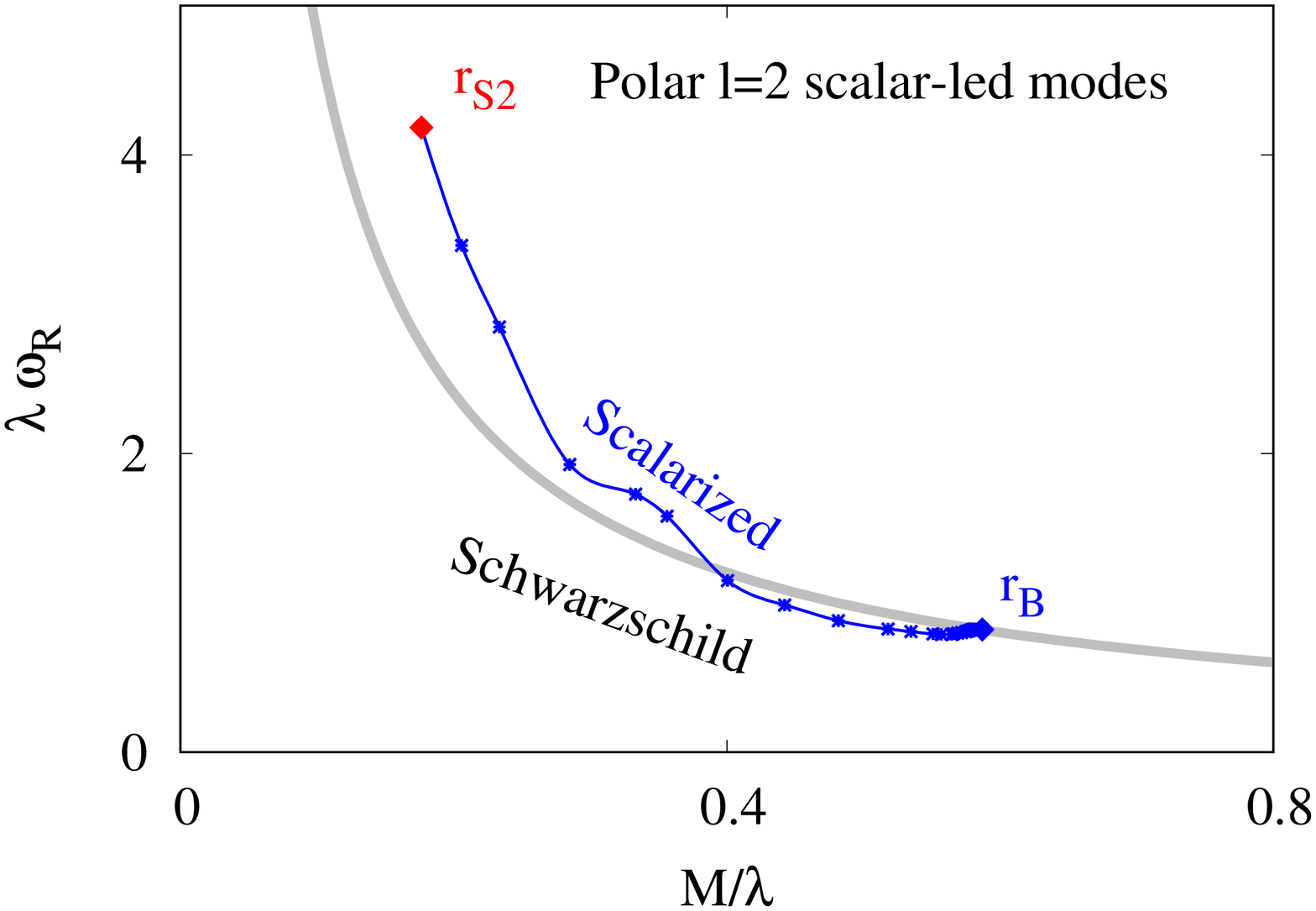}
	\includegraphics[width=0.34\linewidth,angle=-90]{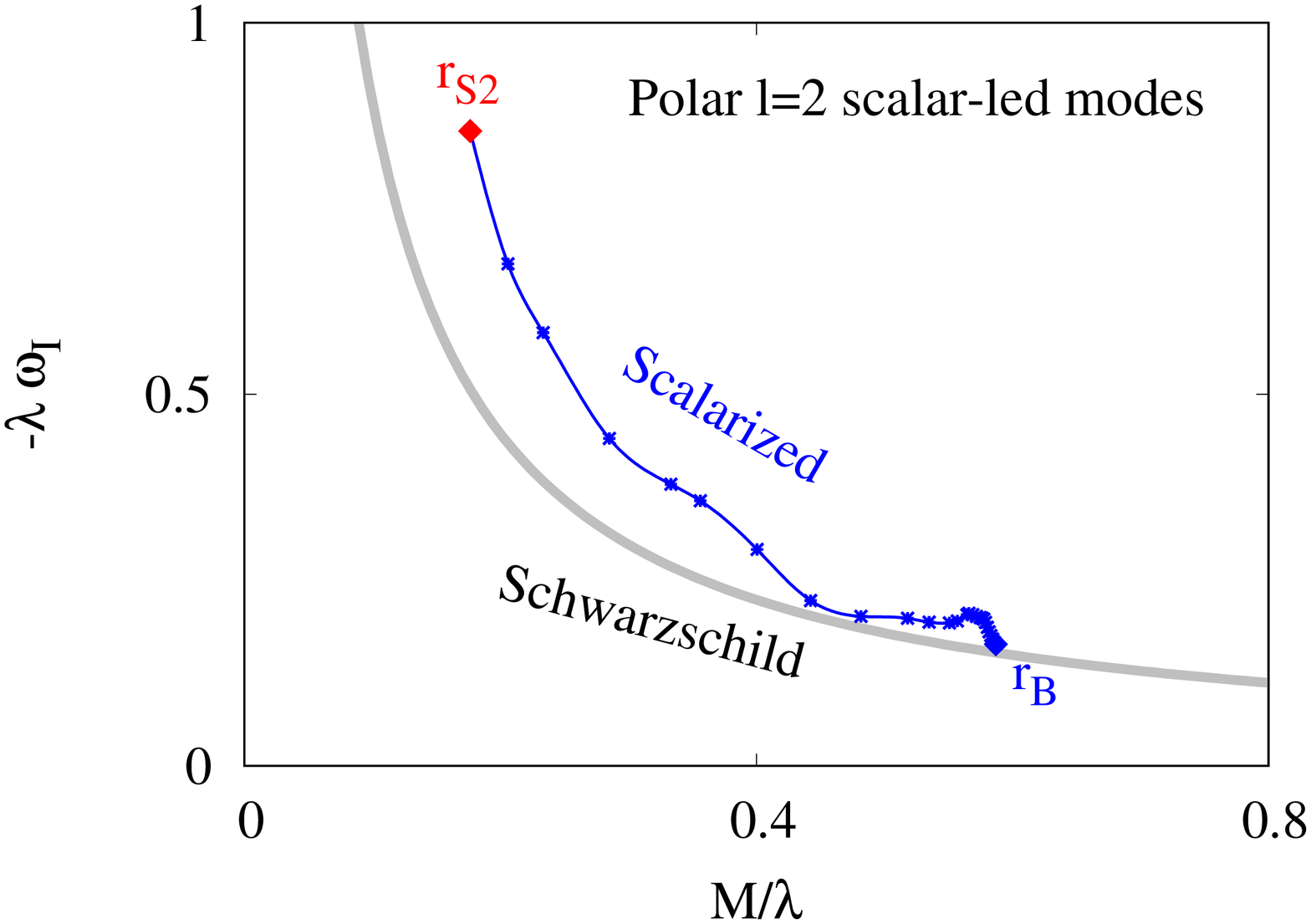}
	\caption{
Scaled polar scalar-led $l=2$ eigenvalue $\lambda \omega$
vs. scaled total mass $M/\lambda$: real part/frequency $\omega_R$
(\textit{left}) and imaginary part/inverse damping time $\omega_I$ (\textit{right}),
depicted in the range $r_{\rm S2} < r_{\rm H}< r_{\rm B}$,
where hyperbolicity is lost at $r_{\rm S2}$, and
$r_{\rm B}$ is the bifurcation point from the Schwarzschild solution.
For comparison also the Schwarzschild mode is shown (solid grey).}
	\label{fig:plotl2_scalar}
\end{figure}
\begin{figure}
	\centering
	\includegraphics[width=0.34\linewidth,angle=-90]{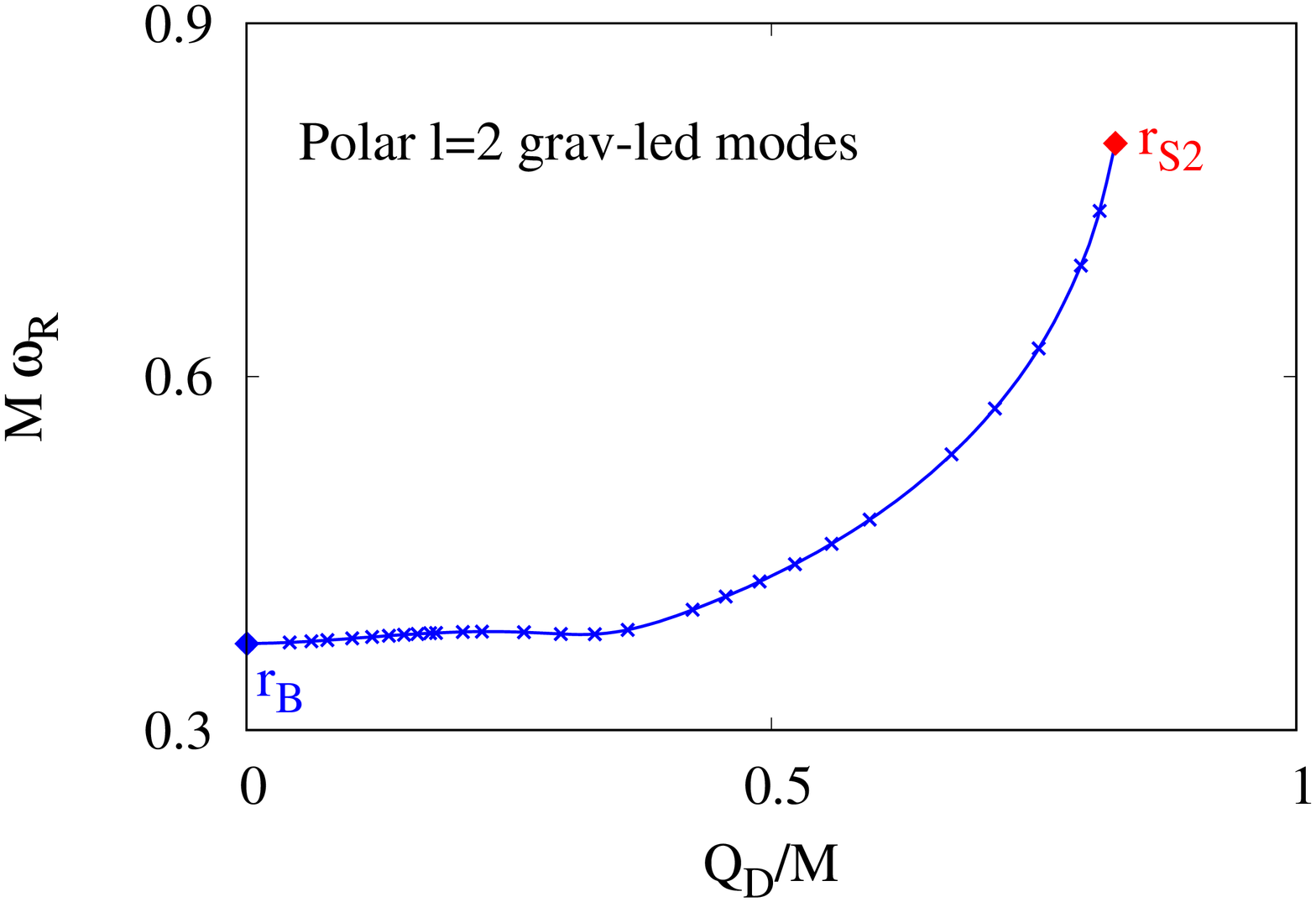}
	\includegraphics[width=0.34\linewidth,angle=-90]{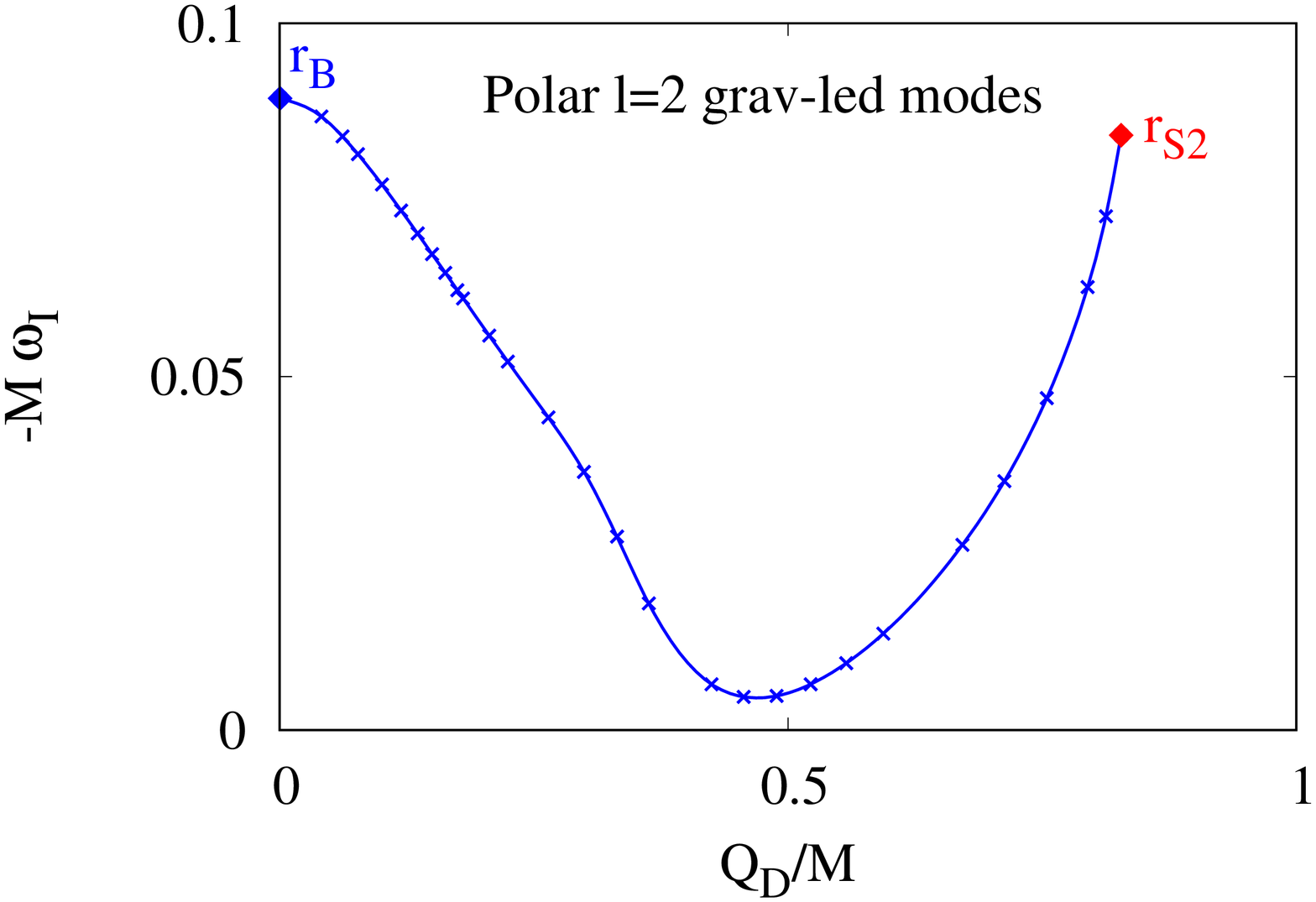}
	\caption{
Scaled polar grav-led $l=2$ eigenvalue $M \omega$
vs. scalar charge $Q_D/M$: real part/frequency $\omega_R$
(\textit{left}) and imaginary part/inverse damping time $\omega_I$ (\textit{right}),
depicted in the range $r_{\rm S2} < r_{\rm H}< r_{\rm B}$.
Hyperbolicity is lost at $r_H=r_{\rm S2}$, the maximum value of $Q_D/M$ shown.
$Q_D/M=0$ corresponds to the bifurcation point $r_{\rm B}$
from the Schwarzschild solution.}
	\label{fig:plotl2_grav_Qd}
\end{figure}
\begin{figure}
	\centering
	\includegraphics[width=0.34\linewidth,angle=-90]{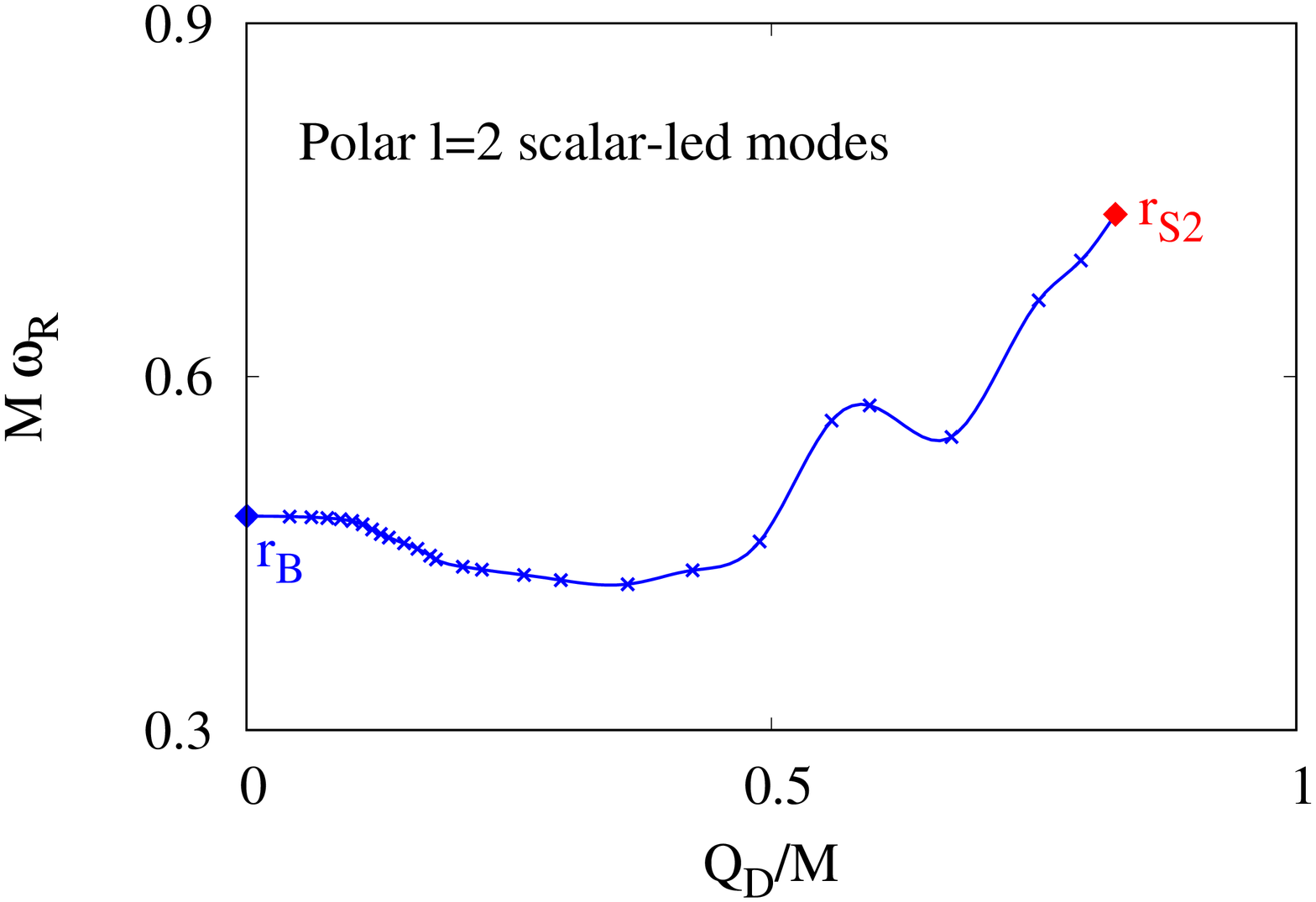}
	\includegraphics[width=0.34\linewidth,angle=-90]{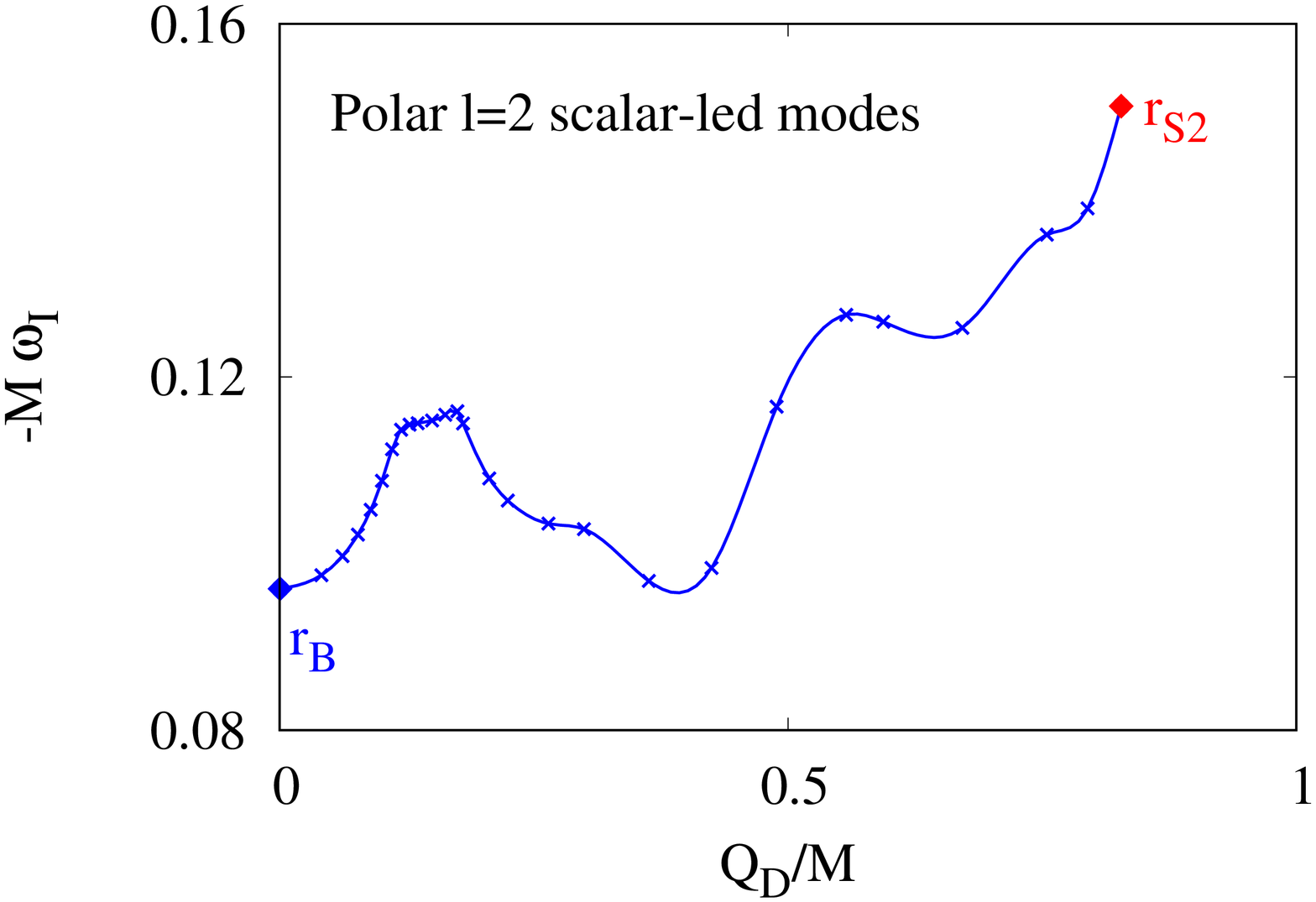}
	\caption{
Scaled polar scalar-led $l=2$ eigenvalue $M \omega$
vs. scalar charge $Q_D/M$: real part/frequency $\omega_R$
(\textit{left}) and imaginary part/inverse damping time $\omega_I$ (\textit{right}),
depicted in the range $r_{\rm S2} < r_{\rm H}< r_{\rm B}$.
Hyperbolicity is lost at $r_H=r_{\rm S2}$, the maximum value of $Q_D/M$ shown.
$Q_D/M=0$ corresponds to the bifurcation point $r_{\rm B}$
from the Schwarzschild solution.}
	\label{fig:plotl2_scalar_Qd}
\end{figure}

In the case of the quadrupole ($l=2$) modes we now have to distinguish between the
grav-led modes and the scalar-led modes, since quadrupole modes
are also present in the absence of a scalar field.
For both families of modes we start again
at the  bifurcation point $r_{\rm H} = r_{\rm B}$,
where the modes of the Schwarzschild solution are
$M\omega=0.3737-i0.08895$ and $M\omega=0.481-i0.0894$
for the gravitational mode and the scalar mode, respectively.
Then we track these modes again along the fundamental branch.

The grav-led $l=2$ mode and the scalar-led $l=2$ mode
are exhibited in Fig.~\ref{fig:plotl2_grav} and Fig.~\ref{fig:plotl2_scalar}, respectively,
where again $\lambda \omega$ is shown versus $M/\lambda$,
with frequency $\omega_R$ on the left and damping rate $\omega_I$ on the right.
Again, the Schwarzschild modes are also shown (solid grey).
The frequency of the grav-led modes is larger than the frequency
of the Schwarzschild modes, with the difference increasing
along the branch. The damping rate, in contrast, is smaller
than the Schwarzschild damping rate,
becoming even very small in some intermediate range, while increasing again towards the
Schwarzschild value as the critical point $r_{\rm S2}$ is approached.

The scalar-led mode exhibits less strong deviations from Schwarzschild.
The frequency is first slightly smaller than the Schwarzschild value,
but then starts to increase more rapidly towards smaller $r_{\rm H}$.
The damping rate, on the other hand, is always
larger than the Schwarzschild value.
Fig.~\ref{fig:plotl2_grav_Qd} and Fig.~\ref{fig:plotl2_scalar_Qd}
show the frequency and damping time of the grav-led and
scalar-led mode again versus the scalar charge.
As before, these diagrams enhance the deviations from Schwarzschild.
Also for the $l=2$ modes, we emphasize that we do not find unstable modes.

\subsection{Isospectrality breaking}

\begin{figure}
	\centering
	\includegraphics[width=0.34\linewidth,angle=-90]{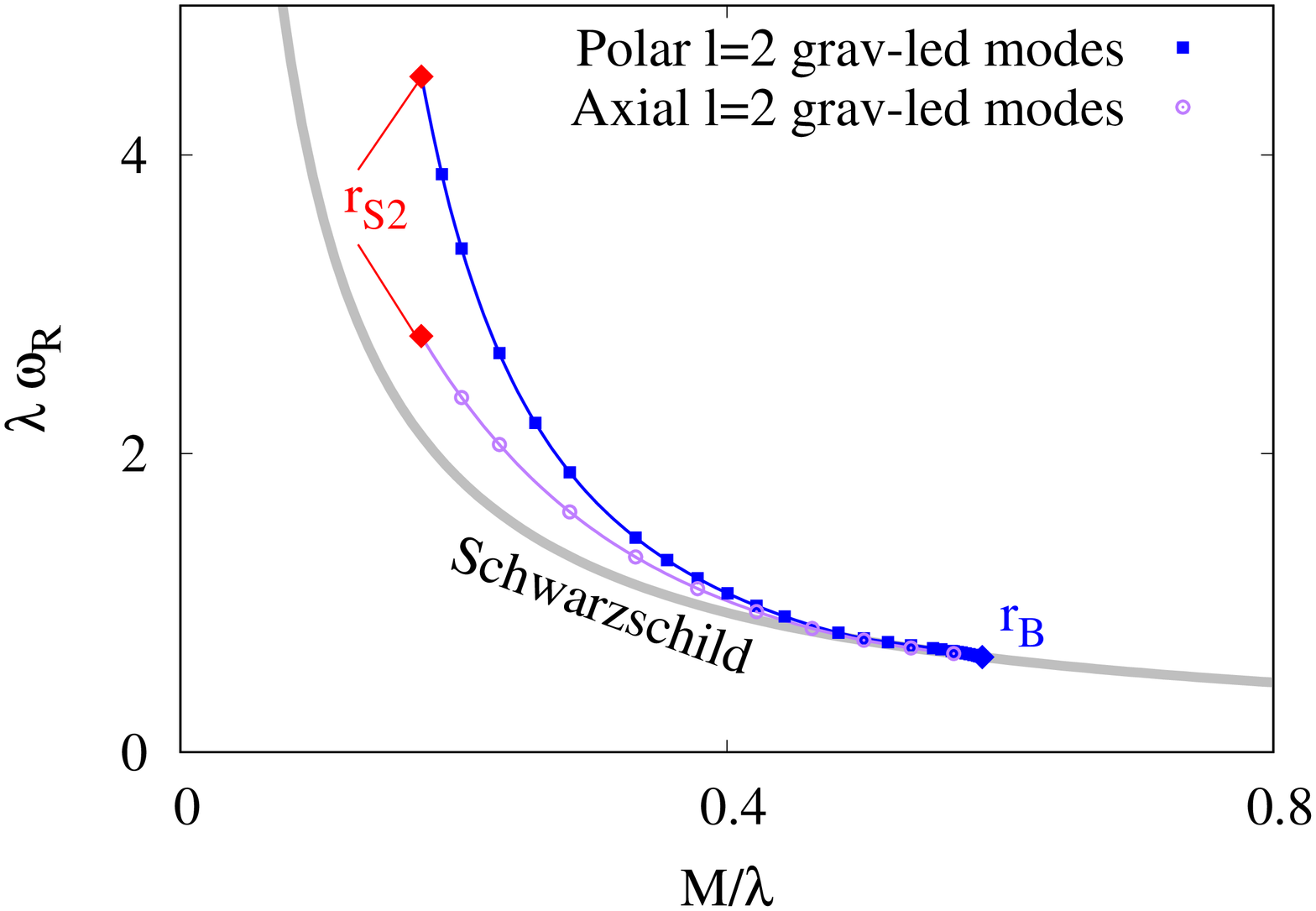}
	\includegraphics[width=0.34\linewidth,angle=-90]{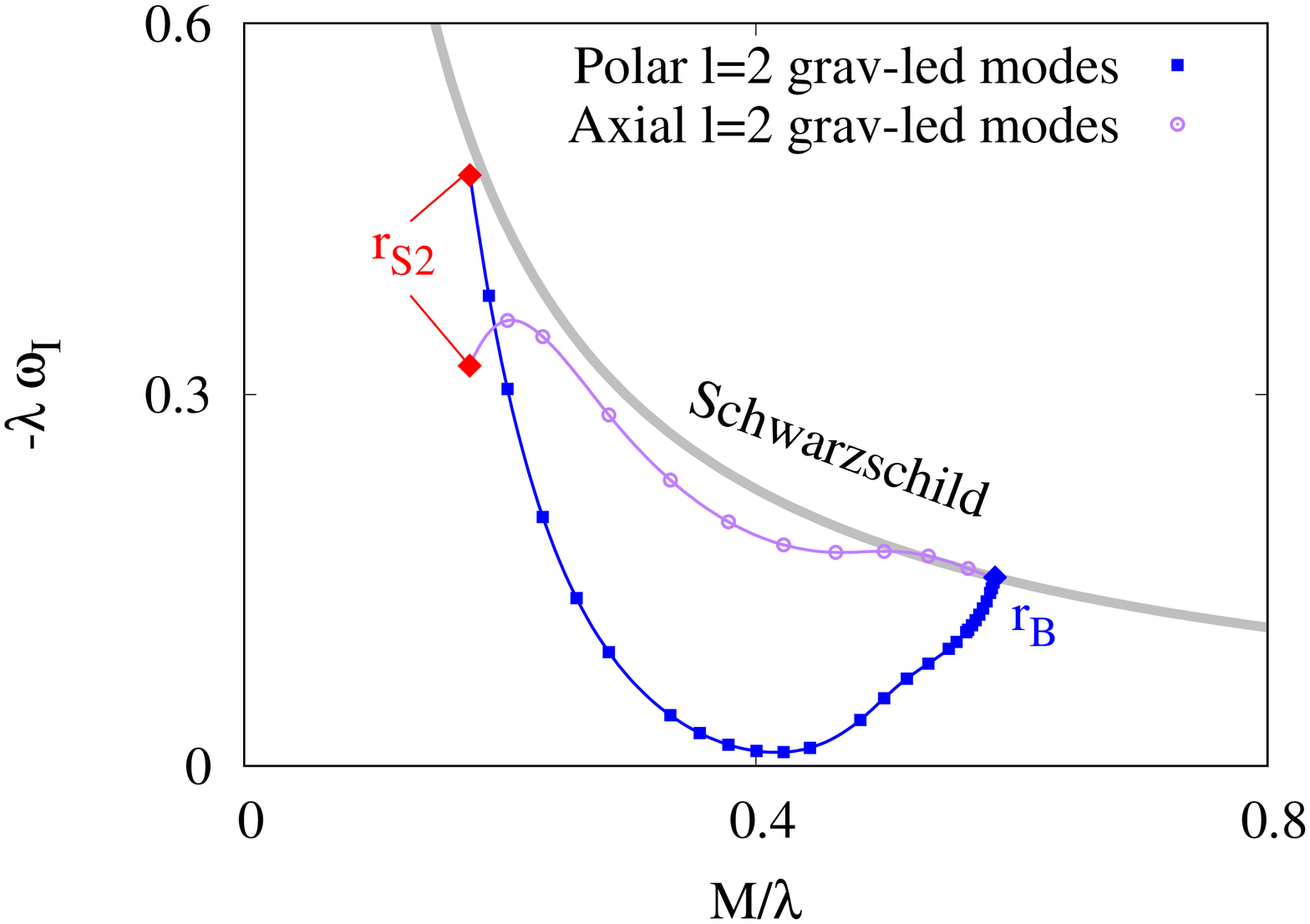}
	\caption{
Scaled $l=2$ eigenvalue $\lambda \omega$
vs. scaled total mass $M/\lambda$
for the polar grav-led mode (blue),
and the axial mode (red): real part/frequency $\omega_R$
(\textit{left}) and imaginary part/inverse damping time $\omega_I$ (\textit{right}),
depicted in the range $r_{\rm S2} < r_{\rm H}< r_{\rm B}$,
where hyperbolicity is lost at $r_{\rm S2}$, and
$r_{\rm B}$ is the bifurcation point from the Schwarzschild solution.
For comparison also the Schwarzschild mode is shown (solid grey).}
	\label{fig:plotl2_iso}
\end{figure}

The modes of the Schwarzschild black holes are known to exhibit
isospectrality,
meaning that the polar and the axial mode spectra coincide.
Previously it has been observed, that isospectrality
is broken in the presence of a dilatonic coupling \cite{Blazquez-Salcedo:2016enn}.
This is not surprising,
since the scalar field enters the polar equations, while it does not
enter the axial equations.
Therefore breaking of isospectrality is also expected for EsGB
black holes. 

We demonstrate the breaking of isospectrality in Fig.~\ref{fig:plotl2_iso},
where we show the modes $\lambda \omega$ versus $M/\lambda$
(with frequency $\omega_R$ on the left and damping rate $\omega_I$ on the right),
comparing the polar grav-led mode (blue),
with the axial mode (red) and also with
the isospectral Schwarzschild modes (solid grey).
We note, that both the polar grav-led and axial frequencies follow the behavior
of the Schwarzschild frequency, but exceed it the more, {the smaller the black hole is},
with the polar mode {frequency} increasing faster with decreasing black hole size.
The  decay rate, on the other hand,  follows for the axial case the Schwarzschild decay rate
to some extent, whereas the polar case features a completely different decay rate,
deviating significantly from the Schwarzschild decay rate except at the boundaries
of the relevant range $r_{\rm S2} < r_{\rm H}< r_{\rm B}$.

\section{Conclusions}

We have studied the polar quasinormal modes on the fundamental branch
of spontaneously scalarized black hole solutions of the EsGB  theory
proposed and studied in \cite{Doneva:2017bvd}.
These scalarized black holes are thermodynamically {preferred over the GR one}s.
However, they should also be dynamically stable
or at least sufficiently long-lived
to be of astrophysical interest. Here the analysis
of their quasinormal modes is an important tool,
since it reveals either linear (mode) stability or instability
of the black holes, and in case {that} the black holes are unstable the analysis
also provides the time-scale of the instability.

The quasinormal modes consist of polar and axial modes, where the
polar modes are composed of radial and non-radial modes.
The radial modes were investigated before,
where the analysis was focused on the detection of potential instabilities
\cite{Blazquez-Salcedo:2018jnn}.
{This} analysis revealed the loss of hyperbolicity of
the perturbation equations at a critical upper size $r_{\rm S1}$
(for a given coupling constant) for the
scalarized black holes. However, for scalarized black holes
with sizes in the range between this critical value
and the bifurcation point $r_{\rm B}$ from the Schwarzschild branch,
no instabilities were detected,
making scalarized black holes in this range
potentially stable physically interesting objects.

Subsequent investigation of the axial modes of these black holes
revealed the loss of hyperbolicity of
the perturbation equations at a critical upper size $r_{\rm S2}$,
slightly larger than $r_{\rm S1}$.
Again, no instabilities were detected in the range
$r_{\rm S2} < r_{\rm H} < r_{\rm B}$,
retaining a slightly smaller interval
of potentially stable physically interesting scalarized black holes
 \cite{Blazquez-Salcedo:2020rhf}.

In the present study we have investigated the non-radial polar modes
of the black holes in this interval $r_{\rm S2} < r_{\rm H} < r_{\rm B}$.
We have calculated the monopole ($l=0$) modes,
the dipole ($l=1$) modes and the quadrupole ($l=2$) modes,
where the quadrupole modes consist of grav-led and scalar-led modes.
Since we did not find any instabilities, we conclude
that the scalarized black holes on  the fundamental branch
are linearly (mode) stable in the range $r_{\rm S2} < r_{\rm H} < r_{\rm B}$
(for fixed coupling).
We note, that we have also continued the analysis of the modes
into the range below $r_{\rm S2}$, where we did not find
any further pathologies.

We have compared the frequencies and the damping times of the
modes of the scalarized black hole with their Schwarzschild counterparts.
We have not seen a general trend for the deviations of the various
multipolar modes, which all feature more or less substantial deviations
in the frequencies and damping times.
An expected outcome of our investigations has been the breaking of isospectrality,
i.e., of the degeneracy of the axial and polar Schwarzschild modes.
Indeed, in the EsGB black holes we have observed a splitting between the axial modes
and the polar grav-led modes as soon as scalarization arises.

This can be relevant with regard to gravitational wave astronomy. With the improvement of sensitivities of the detectors, it could be possible for the LIGO-VIRGO collaboration to measure the ringdown frequencies. In this paper we have quantitatively explored the smoking gun that characterizes the ringdown phase of scalarized black holes in EsGB theory as compared to GR: the richer quasinormal mode spectrum that appears, first, because of the breaking of isospectrality between the grav-led modes in the axial and polar channels, and second, because of the existence of scalar-led modes, that appear at any multipole, and in particular, at $l=0$. Of course, only stable solutions describing astrophysically conceivable objects are relevant candidates in this respect.

While our study has concluded our linear stability analysis, a non-linear
stability analysis would be a natural next objective.
On the other hand, investigation of the instabilities of
rotating spontaneously scalarized black holes in this theory should also be
an  interesting and challenging step forward.

\section*{Acknowledgements}
JLBS, SK, JK and PN gratefully acknowledge support by the DFG funded
Research Training Group 1620 ``Models of Gravity''.
JLBS would like to acknowledge support from the DFG project BL 1553.
DD acknowledges financial support via an Emmy Noether Research Group
funded by the German Research Foundation (DFG) under grant no. DO 1771/1-1.
DD is indebted to the Baden-W\"urttemberg Stiftung for the financial support
of this research project by the Eliteprogramme for Postdocs.
SY would like to thank the University of T\"ubingen for the financial support
and acknowledges the partial support by the Bulgarian NSF Grant KP-06-H28/7.
P.N. is partially supported by the Bulgarian NSF Grant  KP-06-H38/2.
The authors would also like to acknowledge networking support by the
COST Actions CA16104 and CA15117.

\section*{Appendix}

In this appendix we will present the perturbation equations. Using for the metric the Ansatz (\ref{eq:pert_metric}) and for the scalar {field} the Ansatz (\ref{eq:pert_scalar}), we get a system of equations for { the polar parity perturbation functions $H_0,\,H_1,\,H_2,\,T$ and $ \varphi_1$}. For simplicity we introduce the function $K(r)=\frac{1}{1-2m(r)/r}$. Also in the following equations the prime symbol denotes the derivative with respect to $r$: $K'=\frac{dK}{dr}$, etc...

We start with the Einstein equations.
From the $(t,t)$-component we get
\begin{eqnarray}
\label{pert_tt}
	&{T''}&=\,{\left(\frac {4\left( K-1 \right) {\lambda}^
			{2}}{r \left( -rK  {{ e}^{{6{ \varphi_0}}^{2}}}
			+4\,{\lambda}^{2}{ \varphi_0'}\,{ \varphi_0} \right) }\right) \varphi_0\varphi_{1}''}+\,{\left(\frac {
			-4\,K{ \varphi_0}\,{ \varphi_0'}\, \left( K-3 \right) {\lambda}^{2}-2
			\,{K}^{2} {{ e}^{{6{ \varphi_0}}^{2}}} r  {
		}}{2rK \left( -rK {{ e}^{{6{ \varphi_0}}^{2}}} +4\,{\lambda}^{2}{ \varphi_0'}\,{ \varphi_0} \right) }\right) H_{2}''} \nonumber \\
	&+&\,{\left( \frac { \left(
			-8\,K{ \varphi_0}\,{ \varphi_0''}\,r+12\,{ \varphi_0}\,{ \varphi_0'}\,{ K'}
			\,r+8\,K{ \varphi_0'}\, \left( 12\,{{ \varphi_0}}^{2}{ \varphi_0'}\,r-{ \varphi_0'}\,r
			-2\,{ \varphi_0} \right)  \right) {\lambda}^{2}}{2rK \left( -r
			K {{ e}^{{6{ \varphi_0}}^{2}}}+4\,{\lambda}^{2}{
				\varphi_0'}\,{ \varphi_0} \right)} \right. } \nonumber \\
	&+&{\left. \frac{-K{ K'}\,{
				{ e}^{{6{ \varphi_0}}^{2}}}{r}^{2}+6\,{K}^{2} {
				{ e}^{{6{ \varphi_0}}^{2}}}r }{2rK \left( -r
			K {{ e}^{{6{ \varphi_0}}^{2}}}+4\,{\lambda}^{2}{
				\varphi_0'}\,{ \varphi_0} \right) } \right) T'} \nonumber \\
	&+&\,{\left(\frac { \left(  -4\,{\lambda}^{2}{ \varphi_0}\,
			\left( K-3 \right) { K'}-16\,K{ \varphi_0'}\, \left( 12\,{{ \varphi_0}}^{2
			}-1 \right)  \left( K-1 \right)  \right) {\lambda}^{2}+4\,{K}^{2}{
				\varphi_0'}\, {{ e}^{{6{ \varphi_0}}^{2}}}{r}^{2}  }{2rK \left( -rK {{ e}^{{6{ \varphi_0}}^{2}}}
			+4\,{\lambda}^{2}{ \varphi_0'}\,{ \varphi_0} \right) }\right) \varphi_{1}'} \nonumber \\
	&+& {\left(\frac {4\,{\lambda}^{2} \left( K-2 \right) \varphi_0\varphi_{0}''}{r \left( rK
			{{ e}^{{6{ \varphi_0}}^{2}}}-4\,{\lambda}^{2}{
				\varphi_0'}\,{ \varphi_0} \right) }\right. }-{\frac {  \left(rK {{ e}^{{6{ \varphi_0
					}}^{2}}} +2\,K{\lambda}^{2}{ \varphi_0'}\,{ \varphi_0}-12\,{
				\lambda}^{2}{ \varphi_0'}\,{ \varphi_0}\right) K' }{rK \left( rK
			{{ e}^{{6{ \varphi_0}}^{2}}}-4\,{\lambda}^{2}{ \varphi_0'}\,{
				\varphi_0} \right) } } \nonumber \\
	&+&\, { \frac {2\,K{{ \varphi_0'}}^{2} {{ e}^{{6
						{ \varphi_0}}^{2}}} {r}^{3}+{K}^{2} {{ e}^{{6{ \varphi_0}
					}^{2}}} {l}^{2}r+{K}^{2} {{ e}^{{6{ \varphi_0}}^{2}}}
			lr-96\,K{{ \varphi_0}}^{2}{{ \varphi_0'}}^{2}{\lambda}^{2}r+2\,rK
			{{ e}^{{6{ \varphi_0}}^{2}}} -4\,K{ \varphi_0}\,{ \varphi_0'
			}\,{l}^{2}{\lambda}^{2}} {2{r}^{2} \left( rK {
				{ e}^{{6{ \varphi_0}}^{2}}} -4\,{\lambda}^{2}{ \varphi_0'}\,{
				\varphi_0} \right) }  } \nonumber \\
	&+& \left. \frac{192\,{{ \varphi_0}}^{2}{{ \varphi_0'}}^{2}{\lambda}^{2
		}r-4\,K{ \varphi_0}\,{ \varphi_0'}\,l{\lambda}^{2}+8\,K{{ \varphi_0'}}^{2}{\lambda
		}^{2}r-16\,{{ \varphi_0'}}^{2}{\lambda}^{2}r}{2{r}^{2} \left( rK {
			{ e}^{{6{ \varphi_0}}^{2}}} -4\,{\lambda}^{2}{ \varphi_0'}\,{
			\varphi_0} \right)} \right) H_2 \nonumber \\
	&+& \left(\frac { \left( 8\,K{
			\varphi_0}\, \left( K-1 \right) { \varphi_0''}-4\,{ \varphi_0}\,{ \varphi_0'}\,
		\left( K-3 \right) { K'}-8\,K{{ \varphi_0'}}^{2} \left( 12\,{{ \varphi_0}}
		^{2}-1 \right)  \left( K-1 \right)  \right) {\lambda}^{2}}{2rK \left( -rK {{ e}^{{6{ \varphi_0
				}}^{2}}}+4\,{\lambda}^{2}{ \varphi_0'}\,{ \varphi_0} \right)} \right.  \nonumber \\
	&-&\left.\frac{2\,K{ K'}
		\,  {{ e}^{{6{ \varphi_0}}^{2}}}r+2\,{K}^{2}  {
			{ e}^{{6{ \varphi_0}}^{2}}} \left( -{{ \varphi_0'}}^{2}{r}^{2}+
		K-1 \right)  }{2rK \left( -rK {{ e}^{{6{ \varphi_0
				}}^{2}}}+4\,{\lambda}^{2}{ \varphi_0'}\,{ \varphi_0} \right)}\right) H_0 \nonumber \\
	&+&\left( -\,{\frac {2K{ \varphi_0}\,{\lambda}^{2} \left( l+2 \right)
			\left( l-1 \right) { \varphi_0''}}{r \left( rK {{ e}^{{6{ \varphi_0}
					}^{2}}} -4\,{\lambda}^{2}{ \varphi_0'}\,{ \varphi_0} \right) }}+{
		\frac {{\lambda}^{2}{ \varphi_0'}\,{ \varphi_0}\, \left( l+2 \right)  \left( l
			-1 \right) { K'}}{r \left( rK {{ e}^{{6{ \varphi_0}}^{2}}}
			-4\,{\lambda}^{2}{ \varphi_0'}\,{ \varphi_0} \right) }}\right) T \nonumber \\
	&+& \left( \frac {K \left( K {{ e}^{{6{ \varphi_0}}^{2}}}+48\,{
			{ \varphi_0}}^{2}{{ \varphi_0'}}^{2}{\lambda}^{2}-4\,{{ \varphi_0'}}^{2}{\lambda}^
		{2} \right)  \left( l+2 \right)  \left( l-1 \right)}{2r \left( rK
		{{ e}^{{6{ \varphi_0}}^{2}}}-4\,{\lambda}^{2}{
			\varphi_0'}\,{ \varphi_0} \right) }\right)  T \nonumber \\
	&+& \left(\,{\frac {4{\lambda}^{2}
			\left( 12\,{{ \varphi_0}}^{2}-1 \right)  \left( K-1 \right) { \varphi_0''}}{r
			\left( rK  {{ e}^{{6{ \varphi_0}}^{2}}}-4\,{\lambda}
			^{2}{ \varphi_0'}\,{ \varphi_0} \right) }}\right) \varphi_1 \nonumber \\
	&-&\,\left({\frac {2{\lambda}^{2} \left( 12
			\,K{{ \varphi_0}}^{2}{ \varphi_0'}\,r-K{ \varphi_0}\,l(l+1)-36\,{{ \varphi_0}}^{2}{
				\varphi_0'}\,r-K{ \varphi_0'}\,r+3\,{ \varphi_0'}\,r \right) { K'
		}}{K{r}^{2} \left( rK {{ e}^{{6{ \varphi_0}}^{2}}}-4
			\,{\lambda}^{2}{ \varphi_0'}\,{ \varphi_0} \right) }}\right) \varphi_1 \nonumber \\
	&-&\left(\,{\frac {144{ \varphi_0}\,{
				{ \varphi_0'}}^{2}{\lambda}^{2} \left( 2\,{ \varphi_0}-1 \right)  \left( 2\,{
				\varphi_0}+1 \right)  \left( K-1 \right) }{r \left( rK  {{ e}^{6{
						{ \varphi_0}}^{2}}} -4\,{\lambda}^{2}{ \varphi_0'}\,{ \varphi_0}
			\right) }} \right) { \varphi_1} 
\end{eqnarray}

From the $(t,r)$-component we get
\begin{eqnarray}
&{ T'}&=\,{\left(\frac {4\, \left( K-1 \right) {\lambda}^{
			2}}{r \left( -rK  {{ e}^{{6{  \varphi_0}}^{2}}}+4\,{\lambda}^{2}{  \varphi_{0}'}\,{  \varphi_0} \right) } \right) \varphi_{0} \varphi_{1}'}+\,{\left(\frac {
		-4\,f{  \varphi_0}\,{  \varphi_{0}'}\, \left( K-3 \right) {\lambda}^{2}-2
		\,  {{ e}^{{6{  \varphi_0}}^{2}}} fKr
	}{2rf \left( -rK  {{ e}^{{6{  \varphi_0}}^{2}}} +4\,{
			\lambda}^{2}{  \varphi_{0}'}\,{  \varphi_0} \right) }\right) H_2} \nonumber \\
&+&\,\left({\frac {2\,  {{ e}^{{6{  \varphi_0}}^{2}}} K{{
				 \varphi_{0}'}}^{2}{r}^{3}+  {{ e}^{{6{  \varphi_0}}^{2}}}{K}^{2
		}(l+2)(l-1))r-2\,
		{{ e}^{{6{  \varphi_0}}^{2}}}{K}^{2}r+2\,rK  {
			{ e}^{{6{  \varphi_0}}^{2}}} - 96K{{  \varphi_0}}^{2}{{  \varphi_{0}'}}^
		{2}{\lambda}^{2}r }{2K{r}^{2} \left( rK  {{ e}^{{6{  \varphi_0}}^{
					2}}} -4\,{\lambda}^{2}{  \varphi_{0}'}\,{  \varphi_0} \right) }
} \right. \nonumber \\
&+& \,{\frac {96\,{{
				 \varphi_0}}^{2}{{  \varphi_{0}'}}^{2}{\lambda}^{2}r+8\,{ \varphi_{0}''}\,(K-1){  \varphi_0}\,{
			\lambda}^{2}r-4\,K{  \varphi_0}\,{  \varphi_{0}'}\,{\lambda}^{2}l(l+1)-8\,{{  \varphi_{0}'
		}}^{2}{\lambda}^{2}r(1-K)}{2K{r}^{2} \left( rK  {{ e}^{{6{  \varphi_0}}^{
					2}}} -4\,{\lambda}^{2}{  \varphi_{0}'}\,{  \varphi_0} \right)}}  \nonumber \\
&-&\left. {\frac { \left( rK {{ e}^{{6{  \varphi_0}}^{2}}} +2\,K{\lambda}^{2}{  \varphi_{0}'}\,{  \varphi_0}-6\,{\lambda}^{2}{  \varphi_{0}'}\,{
			 \varphi_0} \right) { K'}}{r \left( rK  {{ e}^{{6{  \varphi_0}}^{2}}
		} -4\,{\lambda}^{2}{  \varphi_{0}'}\,{  \varphi_0} \right) {K}^{2}}} \right) \frac{H_{1}}{\omega} \nonumber \\
&+&\,{\left(\frac {  -4\, \left( K-1
		\right)  \left( 24\,f{{  \varphi_0}}^{2}{  \varphi_{0}'}+{ f'}\,{  \varphi_0}-2\,f{
			 \varphi_{0}'} \right) {\lambda}^{2}+4\,  {{ e}^{{6{  \varphi_0}}^{2}}}
		fK{  \varphi_{0}'}\,{r}^{2} }{2rf \left( -rK
		{{ e}^{{6{  \varphi_0}}^{2}}} +4\,{\lambda}^{2}{
			 \varphi_{0}'}\,{  \varphi_0} \right) } \right)  \varphi_1} \nonumber \\
&+& { \left( \frac {  4\,{  \varphi_0}\,{  \varphi_{0}'}\, \left( {
			f'}\,r-2\,f \right) {\lambda}^{2}-rK  {{ e}^{{6{  \varphi_0}}^{2}}}
		\left( { f'}\,r-2\,f \right)  }{2rf \left( -r
		K  {{ e}^{{6{  \varphi_0}}^{2}}} +4\,{\lambda}^{2}{
			 \varphi_{0}'}\,{  \varphi_0} \right) } \right) T} \, .
\end{eqnarray}

From the $(t, \varphi)$-component we get
\begin{eqnarray}
\label{pert_tphi}
{H_1'}&=&K\omega H_2 -{\frac {4{\lambda}^{2}{ \varphi_0}{K'}}{4 {\lambda}^{2}{ \varphi_0}{ \varphi_0'}-rK {{e}^{6{{ \varphi_0}}^{2}}}}}\omega \varphi_1 -{\frac { -{f'}  {{e}^{6{{ \varphi_0}}^{2}}}{K}^{2}{r}^{2}+{K'}  {{e}^{6{{ \varphi_0}}^{2}}}f{r}^{2}K }{2f Kr \left(4 {\lambda}^{2}{ \varphi_0}{ \varphi_0'}-rK {{e}^{6{{ \varphi_0}}^{2}}} \right)}}{H_1} \nonumber \\
&-&{\frac {\left(2 { \varphi_0} { \varphi_0'} {K'} r -4 K{ \varphi_0} { \varphi_0''} r+4 K{{ \varphi_0'}}^{2}r \left( 12 {{ \varphi_0}}^{2}-1 \right) \right) {\lambda}^{2}+{K}^{2} {{e}^{6{{ \varphi_0}}^{2}}}r}{4 {\lambda}^{2}{ \varphi_0}{ \varphi_0'}-rK {{e}^{6{{ \varphi_0}}^{2}}}}} \omega T \nonumber  \\
&-&{\frac { 4 f K{ \varphi_0} { \varphi_0''} r-6 f{ \varphi_0} { \varphi_0'} {K'} r+2 K{ \varphi_0'}  \left(2 f{ \varphi_0'} r(1-12 {{ \varphi_0}}^{2})+{ \varphi_0}({f'}r-2 f) \right) }{f Kr \left(4 {\lambda}^{2}{ \varphi_0}{ \varphi_0'} -rK {{e}^{6{{ \varphi_0}}^{2}}} \right)}}{\lambda}^{2}{H_1}  \, .
\end{eqnarray}

From the $(r,r)$-component we get
\begin{eqnarray}
\label{pert_rr}
&{ H_{0}'}&= \left({\frac { \left( 12\,{  \varphi_0}\,{  \varphi_{0}'
		}\,{\lambda}^{2}{r}-{{ e}^{6\,{{  \varphi_0}}^{2}}}K{r}^{2} \right)
		{ f'}}{2f \left( 2\,{  \varphi_0}\,{  \varphi_{0}'}\, \left( K-3 \right) {
			\lambda}^{2}+rK{{ e}^{6\,{{  \varphi_0}}^{2}}} \right) }}-{\frac {rK{
			{ e}^{6\,{{  \varphi_0}}^{2}}}}{2\,{  \varphi_0}\,{  \varphi_{0}'}\, \left( K-3
		\right) {\lambda}^{2}+rK{{ e}^{6\,{{  \varphi_0}}^{2}}}}} \right) {
	T'} \nonumber \\
&+& \left( \,{\frac {-2{  \varphi_0}\, \left( K-3 \right) {\lambda}^{2}{
			f'}}{f \left( 2\,{  \varphi_0}\,{  \varphi_{0}'}\, \left( K-3 \right) {\lambda
		}^{2}+rK{{ e}^{6\,{{  \varphi_0}}^{2}}} \right) }}+\,{\frac {2{{ e}^{
				6\,{{  \varphi_0}}^{2}}}K{  \varphi_{0}'}\,{r}^{2}}{2\,{  \varphi_0}\,{  \varphi_{0}'}\,
		\left( K-3 \right) {\lambda}^{2}+rK{{ e}^{6\,{{  \varphi_0}}^{2}}}}}
\right) {  \varphi_{1}'} \nonumber \\
&+& \left( \,{\frac {-6{\lambda}^{2}{ f'}\,{  \varphi_{0}'
		}\,{  \varphi_0}}{f \left( 2\,{  \varphi_0}\,{  \varphi_{0}'}\, \left( K-3 \right) {
			\lambda}^{2}+rK{{ e}^{6\,{{  \varphi_0}}^{2}}} \right) }}+{\frac {{
			{ e}^{6\,{{  \varphi_0}}^{2}}}{K}^{2}}{2\,{  \varphi_0}\,{  \varphi_{0}'}\, \left( K
		-3 \right) {\lambda}^{2}+rK{{ e}^{6\,{{  \varphi_0}}^{2}}}}} \right) {
	H_2} \nonumber \\
&+&\,{\left(\frac {  8\,{  \varphi_0}\,{  \varphi_{0}'}\,r \left( K-3
		\right) {\lambda}^{2}+4\,{{ e}^{6\,{{  \varphi_0}}^{2}}}K{r}^{2}
	}{2rf \left( 2\,{  \varphi_0}\,{  \varphi_{0}'}\, \left( K
		-3 \right) {\lambda}^{2}+rK{{ e}^{6\,{{  \varphi_0}}^{2}}} \right) } \right) \omega H_1} \nonumber \\
&+&\,{\left(\frac { -4\,fK{  \varphi_0}\,{  \varphi_{0}'}\,l \left( l+1 \right) {
			\lambda}^{2}+{{ e}^{6\,{{  \varphi_0}}^{2}}}f{K}^{2}lr \left( l+1
		\right)   }{2rf \left( 2\,{  \varphi_0}\,{  \varphi_{0}'}\, \left(
		K-3 \right) {\lambda}^{2}+rK{{ e}^{6\,{{  \varphi_0}}^{2}}} \right) }\right) H_0} \nonumber \\
&+&\left( -{\frac {K{  \varphi_0}\,{  \varphi_{0}'}\, \left( l+2 \right)  \left( l-1
		\right) {\lambda}^{2}{ f'}}{f \left( 2\,{  \varphi_0}\,{  \varphi_{0}'}\,
		\left( K-3 \right) {\lambda}^{2}+rK{{ e}^{6\,{{  \varphi_0}}^{2}}}
		\right) }}+\,{\frac { \left( 8\,K{  \varphi_0}\,{  \varphi_{0}'}\,{\lambda}^{2
		}{r}^{2}-2\,{{ e}^{6\,{{  \varphi_0}}^{2}}}{K}^{2}{r}^{3} \right) {
			\omega}^{2}}{2rf \left( 2\,{  \varphi_0}\,{  \varphi_{0}'}\, \left( K-3 \right) {
			\lambda}^{2}+rK{{ e}^{6\,{{  \varphi_0}}^{2}}} \right) }} \right. \nonumber \\
&+& \,\left.{\frac {{
			{ e}^{6\,{{  \varphi_0}}^{2}}}{K}^{2} \left( l+2 \right)  \left( l-1
		\right) }{4\,{  \varphi_0}\,{  \varphi_{0}'}\, \left( K-3 \right) {\lambda}^{2}+2r
		K{{ e}^{6\,{{  \varphi_0}}^{2}}}}} \right) T \nonumber \\
&+& \left( \,{\frac {
		\left( 48\,K{{  \varphi_0}}^{2}{  \varphi_{0}'}\,r-4\,K{  \varphi_0}\,{l}^{2}-144\,{{
				 \varphi_0}}^{2}{  \varphi_{0}'}\,r-4\,K{  \varphi_{0}'}\,r-4\,K{  \varphi_0}\,l+12\,{  \varphi_{0}'
		}\,r \right) {\lambda}^{2}{ f'}}{2rf \left( 2\,{  \varphi_0}\,{  \varphi_{0}'}\,
		\left( K-3 \right) {\lambda}^{2}+rK{{ e}^{6\,{{  \varphi_0}}^{2}}}
		\right) }} \right. \nonumber \\
&-&\left. \,{\frac {4K{  \varphi_0}\, \left( K-1 \right) {\lambda}^{2}{
			\omega}^{2}}{f \left( 2\,{  \varphi_0}\,{  \varphi_{0}'}\, \left( K-3 \right) {
			\lambda}^{2}+rK{{ e}^{6\,{{  \varphi_0}}^{2}}} \right) }} \right) {
	 \varphi_1}   \, .
\end{eqnarray}

From the $(r,\theta)$-component we get
\begin{eqnarray}
\label{pert_rtheta}
&{T'}&={\left( \frac {  2\, {{ e}^{{6{  \varphi_0}}
				^{2}}} fK{r}^{2}-8\,f{  \varphi_0}\,{  \varphi_{0}'}\,{\lambda}^{2}r
	}{2{r}^{2} \left( -K {{ e}^{{6{  \varphi_0}}^{2}}
		}f +2\,{\lambda}^{2}{ f'}\,{  \varphi_{0}'}\,{  \varphi_0} \right)
	} \right) H_{0}'}-{\left( \frac {4{ f'}\,{\lambda}^{2}}{r \left( -K
		{{ e}^{{6{  \varphi_0}}^{2}}} f+2\,{\lambda}^{2}{
			f'}\,{  \varphi_{0}'}\,{  \varphi_0} \right) }\right)  \varphi_0 \varphi_{1}' } \nonumber \\
&+& {\left( \frac {  12\,{ f'}
		\,{  \varphi_0}\,{  \varphi_{0}'}\,{\lambda}^{2}r- {{ e}^{{6{  \varphi_0}}^{2}
		}} Kr \left( { f'}\,r+2\,f \right)  }{2{
			r}^{2} \left( -K  {{ e}^{{6{  \varphi_0}}^{2}}} f+2\,{
			\lambda}^{2}{ f'}\,{  \varphi_{0}'}\,{  \varphi_0} \right) } \right) H_2 }+{\left(\frac {
		-2\, \left( {{ e}^{{{  \varphi_0}}^{2}}} \right) ^{6}K{r}^{2}+8
		\,{  \varphi_0}\,{  \varphi_{0}'}\,{\lambda}^{2}r  }{2{r}^{2
		} \left( -K \left( {{ e}^{{{  \varphi_0}}^{2}}} \right) ^{6}f+2\,{
			\lambda}^{2}{ f'}\,{  \varphi_{0}'}\,{  \varphi_0} \right) }\right) \omega H_1} \nonumber \\
&+&{\left( \frac {
		-4\,{  \varphi_0}\,{  \varphi_{0}'}\, \left( { f'}\,r-2\,f \right) {
			\lambda}^{2}+rK \left( {{ e}^{{{  \varphi_0}}^{2}}} \right) ^{6} \left(
		{ f'}\,r-2\,f \right)  }{2{r}^{2} \left( -K  {
			{ e}^{{6{  \varphi_0}}^{2}}} f+2\,{\lambda}^{2}{ f'}\,{
			 \varphi_{0}'}\,{  \varphi_0} \right) } \right) H_0} \nonumber \\
&+&{ \left( \frac {  8\,{ f'}\,
		\left( 12\,{{  \varphi_0}}^{2}{  \varphi_{0}'}\,r-{  \varphi_{0}'}\,r+{  \varphi_0} \right) {
			\lambda}^{2}+8\, {{ e}^{{6{  \varphi_0}}^{2}}} fK{
			 \varphi_{0}'}\,{r}^{2}}{2{r}^{2} \left( -K {{ e}^{{6{
						 \varphi_0}}^{2}}} f+2\,{\lambda}^{2}{ f'}\,{  \varphi_{0}'}\,{
			 \varphi_0} \right) } \right)  \varphi_1}  \, .
\end{eqnarray}

From the $(\theta, \phi)$-component we get
\begin{eqnarray}
\label{pert_tphi}
&{H_2}&={\left(\frac {f \left(  \left( -4\,K{  \varphi_0}\,{ \varphi_0''}+2
		\,{  \varphi_0}\,{ K'}\,{  \varphi_0'}+4\,K{{  \varphi_0'}}^{2} \left( 12\,{{
				 \varphi_0}}^{2}-1 \right)  \right) {\lambda}^{2}+{K}^{2}  {{e}^{6{{
						 \varphi_0}}^{2}}}  \right)}{ \left( -K  {
			{e}^{{6{  \varphi_0}}^{2}}}  f+2\,{\lambda}^{2}{ f'}\,{
			 \varphi_0'}\,{  \varphi_0} \right) K}\right) H_0} \nonumber \\
&+&\,{\left(\frac {2{\lambda}^{2}
		\left( { f'}\,f{ K'}-K \left( 2\,f{ f''}-{{ f'}}^{2}
		\right)  \right) }{ \left( -K  {{e}^{{6{  \varphi_0}}^{2}
		}} f+2\,{\lambda}^{2}{ f'}\,{  \varphi_0'}\,{  \varphi_0}
		\right) fK}\right)  \varphi_0 \varphi_1}  \, .
\end{eqnarray}

Finally, from the scalar field equation we get
\begin{eqnarray}
& \varphi_1''&=
\left( -{\frac {{f'}{ \varphi_0} \left( K-3 \right) {\lambda}^{2}}{2fK{r}^{2} \left( {{e}^{6{{ \varphi_0}}^{2}}}\right)}}+\frac{ \varphi_0'}{2} \right) {H_2'} +{\left(\frac {{ \varphi_0} f\left( K-1 \right){\lambda}^{2}}{fK{r}^{2} \left( {{e}^{{6{ \varphi_0}}^{2}}} \right)}\right)}{H_0''}
-{\left(\frac {{ \varphi_0} {f'}r{\lambda}^{2}}{fK{r}^{2} \left( {{e}^{{6{ \varphi_0}}^{2}}} \right)}\right)} {T''}
\nonumber \\
&+& \left( {\frac {{K'}}{2K}}-{\frac {{f'}r+4f}{2fr}} \right) { \varphi_1'}
+ \left({\frac {Kl \left( l+1\right) }{{r}^{2}}} -{\frac {K{\omega}^{2}}{f}}\right)  \varphi_1
\nonumber \\
&-&{\left(\frac {2{ \varphi_0}\left( K-1 \right) {\lambda}^{2}\omega}{fK{r}^{2} \left( {{e}^{{6{ \varphi_0}}^{2}}} \right)}\right)}{H_1'}+ \left(  \left( -{\frac {{ \varphi_0} \left( K-3 \right) {K'}}{2{r}^{2} \left( {{e}^{{6{ \varphi_0}}^{2}}} \right){K}^{2}}}+{\frac {{f'}{ \varphi_0}\left( K-1 \right) }{fK{r}^{2} \left( {{\rm e}^{{{ \varphi_0}}^{2}}}\right) ^{6}}} \right) {\lambda}^{2}-\frac{ \varphi_0'}{2} \right) {H_0'} \nonumber  \\
&+&\left(  \left( {\frac {3{f'}{ \varphi_0}{K'}}{2fr \left( {{e}^{{6{ \varphi_0}}^{2}}} \right){K}^{2}}}+{\frac {{ \varphi_0} \left( -2{f''}fr+{{f'}}^{2}r-4{f'}f \right) }{2K{r}^{2}{f}^{2} \left( {{e}^{{6{ \varphi_0}}^{2}}} \right)}}\right) {\lambda}^{2}-{ \varphi_0'} \right) {T' } \nonumber  \\
&+&\left( {\frac {{ \varphi_0} \left( K-1 \right) {\lambda}^{2}{\omega}^{2}}{f{r}^{2} \left( {{e}^{{6{ \varphi_0}}^{2}}} \right)}}+{ \varphi_0''}-{\frac {{ \varphi_0'}{K'}}{2K}}+{\frac {{ \varphi_0'} \left( {f'}r+4f \right) }{2fr}} \right) {H_2} \nonumber  \\
&+&\left( \left( {\frac {{f'}{ \varphi_0} \left( K-6 \right) {K'}}{2f{r}^{2}\left( {{e}^{{6{ \varphi_0}}^{2}}} \right){K}^{2}}}+{\frac {{ \varphi_0} \left( {f'}{l(l+1)}fK+(2{f''}fr-{{f'}}^{2} r)(2-K) \right) }{2K{f}^{2} \left( {{e}^{{6{ \varphi_0}}^{2}}} \right){r}^{3}}} \right) {\lambda}^{2} \right) {H_2} \nonumber \\
&+& \left( {\frac {{ \varphi_0} \left( K-3 \right) {K'}{\lambda}^{2}}{f{r}^{2} \left( {{e}^{{6{ \varphi_0}}^{2}}} \right){K}^{2}}}+{\frac {{ \varphi_0'}}{f}} \right) \omega H_1-{\left(\frac {{ \varphi_0}l \left( l+1 \right) {K'}{\lambda}^{2}}{2\left( {{e}^{{6{ \varphi_0}}^{2}}} \right)K{r}^{3}}\right)} H_0 \nonumber  \\
&+& \left( {\frac {{K'}{ \varphi_0}{\lambda}^{2}{\omega}^{2}}{rKf \left( {{e}^{{6{ \varphi_0}}^{2}}} \right)}}+ \left({\frac {{f'}{ \varphi_0} \left( l+2 \right)  \left( 1-l \right) {K'}}{4fK{r}^{2} \left( {{e}^{6{{ \varphi_0}}^{2}}} \right)}}-{\frac{{ \varphi_0} \left({{f'}}^{2}-2{f''}f \right)  \left( l+2\right)  \left( l-1 \right) }{4{r}^{2}{f}^{2} \left( {{e}^{{6{ \varphi_0}}^{2}}} \right)}} \right) {\lambda}^{2} \right)T \nonumber  \\
&+& \left( \left({\frac {{f'} \left( 12{{ \varphi_0}}^{2}-1 \right)  \left( K-3 \right) {K'}}{2f{r}^{2}\left({{e}^{{6{ \varphi_0}}^{2}}} \right){K}^{2}}}+{\frac { \left( 12{{ \varphi_0}}^{2}-1 \right)  \left( K-1 \right)  \left({{f'}}^{2}-2{f''}f \right) }{2K{r}^{2}{f}^{2} \left( {{e}^{{6{ \varphi_0}}^{2}}} \right)}} \right) {\lambda}^{2}\right)  \varphi_1  \, .
\label{pert_scalar}
\end{eqnarray}

The remaining components of the Einstein equations are zero or can be written in terms of the previous components.

{This system of equations can be further simplified. After some algebra we can derive an equation which relates the function $H_0$ to the rest of the metric perturbation functions, as well as the perturbation of the scalar field $\varphi_1$ and its derivative. Using this relation and the algebraic relation $(\ref{pert_tphi})$ we can eliminate the functions $H_0$ and $H_2$, and reduce the perturbation equations to a coupled system of three differential equations for the remaining metric perturbation functions $H_1$ and $T$ and the perturbation of the scalar field $\varphi_1$. It can be proven that these equations together with the two relations representing the functions $H_0$ and $H_2$ in terms of the rest of the perturbation functions are equivalent to the initial set of perturbation equations. }

{The system of equations for the perturbations functions  $H_1$, $T$ and $\varphi_1$ can be represented schematically in the form}
\be
\left(
\begin{array}{c}
 H_1' \\
 T' \\
  \varphi_1' \\
  \varphi_1''
\end{array}\right)
+\left(\begin{array}{c}
V_{11} \;\; V_{12} \;\; V_{13} \;\; V_{14} \\
V_{21}\;\; V_{22}\;\; V_{23} \;\; V_{24} \\
V_{31}\;\; V_{32}\;\; V_{33}\;\; V_{34} \\
V_{41}\;\; V_{42}\;\; V_{43} \;\;V_{44}
\end{array}\right)\left(
\begin{array}{c}
 H_1 \\
 T \\
  \varphi _1\\
  \varphi _1'
\end{array}\right)
=0 ,
\label{eq:expanper}
\ee
{where $V_{3k}=0$ for $k\neq 4$, and $V_{34}=-1$.} 
Due to the complexity of the components of 
$V_p$, we shall not show them here. However, by making the dependence in $\omega$ explicit, we can write the previous equation like
\be
\left(
\begin{array}{c}
	H_1'/\omega \\
	T' \\
	 \varphi _1' \\
	 \varphi _1''
\end{array}\right)
+
\left[
\left(\begin{array}{c}
	V_{11}^{(1)} \;\; V_{12}^{(1)} \;\; V_{13}^{(1)} \;\; V_{14}^{(1)} \\
	V_{21}^{(1)}\;\; V_{22}^{(1)}\;\; V_{23}^{(1)} \;\; V_{24}^{(1)} \\
	0\;\;\;\;\;\; 0\;\;\;\;\;\; 0\;\;\;\;\;\; -1 \\
	V_{41}^{(1)}\;\; V_{42}^{(1)}\;\; V_{43}^{(1)} \;\;V_{44}^{(1)}
\end{array}\right)
+\omega^2\left(\begin{array}{c}
	V_{11}^{(2)} \;\; V_{12}^{(2)} \;\; V_{13}^{(2)} \;\;\; 0 \\
	V_{21}^{(2)}\;\; V_{22}^{(2)}\;\; V_{23}^{(2)} \;\;\; 0 \\
	0\;\;\;\;\;\;\; 0\;\;\;\;\;\;\; 0\;\;\;\;\;\;\; 0 \\
	V_{41}^{(2)}\;\; V_{42}^{(2)}\;\; V_{43}^{(2)} \;\;\;0
\end{array}\right)
\right]
\left(
\begin{array}{c}
	H_1/\omega \\
	T \\
	 \varphi _1\\
	 \varphi _1'
\end{array}\right)
=0 ,
\label{eq:expanper}
\ee
where the functions $V_{ij}^{(1,2)}$ only depend on $f(r)$, $m(r)$, $ \varphi_0(r)$, $\lambda$ and $l$.
The boundary conditions for the perturbations can be found with the help of the expansions of the background metric and the scalar field. Because of the natural divergence at spatial infinity, one has to consider a high-order expansion of the perturbations at infinity for the quasinormal modes.

For the lowest values $l=0,1$, simpler gauge choices can be chosen~\cite{Zerilli:1971wd}. We note that for $l=1$ we can choose a gauge in which $T$ vanishes identically. Therefore, we can use the first two equations in~\eqref{eq:expanper} to eliminate $H_1'$ and $H_1$ in favor of $ \varphi_1$, $ \varphi_1'$. In the case of $l=0$ we can pick a gauge such that both $T$ and $H_1$ vanish. The equations are then again reduced to a second order equation for the scalar perturbation. This is possible since some harmonics in the expansion are identically zero when $l=0$ or $l=1$.

\bibliographystyle{ieeetr}
\bibliography{EGB_polar}

\end{document}